\documentclass[10pt,twocolumn,letterpaper]{article}

\usepackage{helper_pkg}
\usepackage{times}
\usepackage{epsfig}
\usepackage{graphicx}
\usepackage{amsmath}
\usepackage{amssymb}
\usepackage{booktabs}

\usepackage{bm}
\usepackage{caption}
\usepackage{subcaption}
\usepackage{multirow}
\usepackage{diagbox}
\usepackage{nicematrix}
\usepackage{authblk}
\usepackage{appendix}
\usepackage[accsupp]{axessibility}  

\usepackage[style=ieee]{biblatex}
\usepackage[breaklinks=true,bookmarks=false]{hyperref}

\begin{document}

\title{3D Neural Sculpting (3DNS): Editing Neural Signed Distance Functions}

\renewcommand*{\Authsep}{\hspace{2em}}
\renewcommand*{\Authand}{\hspace{2em}}
\renewcommand*{\Authands}{\hspace{2em}}
\renewcommand*{\Affilfont}{\normalsize\normalfont}
\renewcommand*{\Authfont}{\bfseries}

\author[1]{Petros Tzathas}
\author[1]{Petros Maragos}
\author[2,3]{Anastasios Roussos}
\affil[1]{School of Electrical \& Computer Engineering, National Technical University of Athens, Greece}
\affil[2]{Institute of Computer Science (ICS), Foundation for Research \& Technology - Hellas (FORTH), Greece}
\affil[3]{College of Engineering, Mathematics and Physical Sciences, University of Exeter, UK}

\maketitle

\newcommand{\editfigbust}[1]{
    \parbox[c]{1.05in}{
        \includegraphics[trim=450 50 450 20, clip, width=1.05in]{Figures/edit_figures/bust/bust_#1.png}
    }
}

\newcommand{\editfigbunny}[1]{
    \parbox[c]{1.05in}{
        \includegraphics[trim=250 0 250 0, clip, width=1.05in]{Figures/edit_figures/bunny/bunny_#1.png}
    }
}

\newcommand{\editfigpumpkin}[1]{
    \parbox[c]{1.05in}{
        \includegraphics[trim=250 0 250 100, clip, width=1.05in]{Figures/edit_figures/pumpkin/pumpkin_#1.png}
    }
}

\newcommand{\arrowfig}{
    \parbox[c]{0.3in}{
        \includegraphics[width=0.3in]{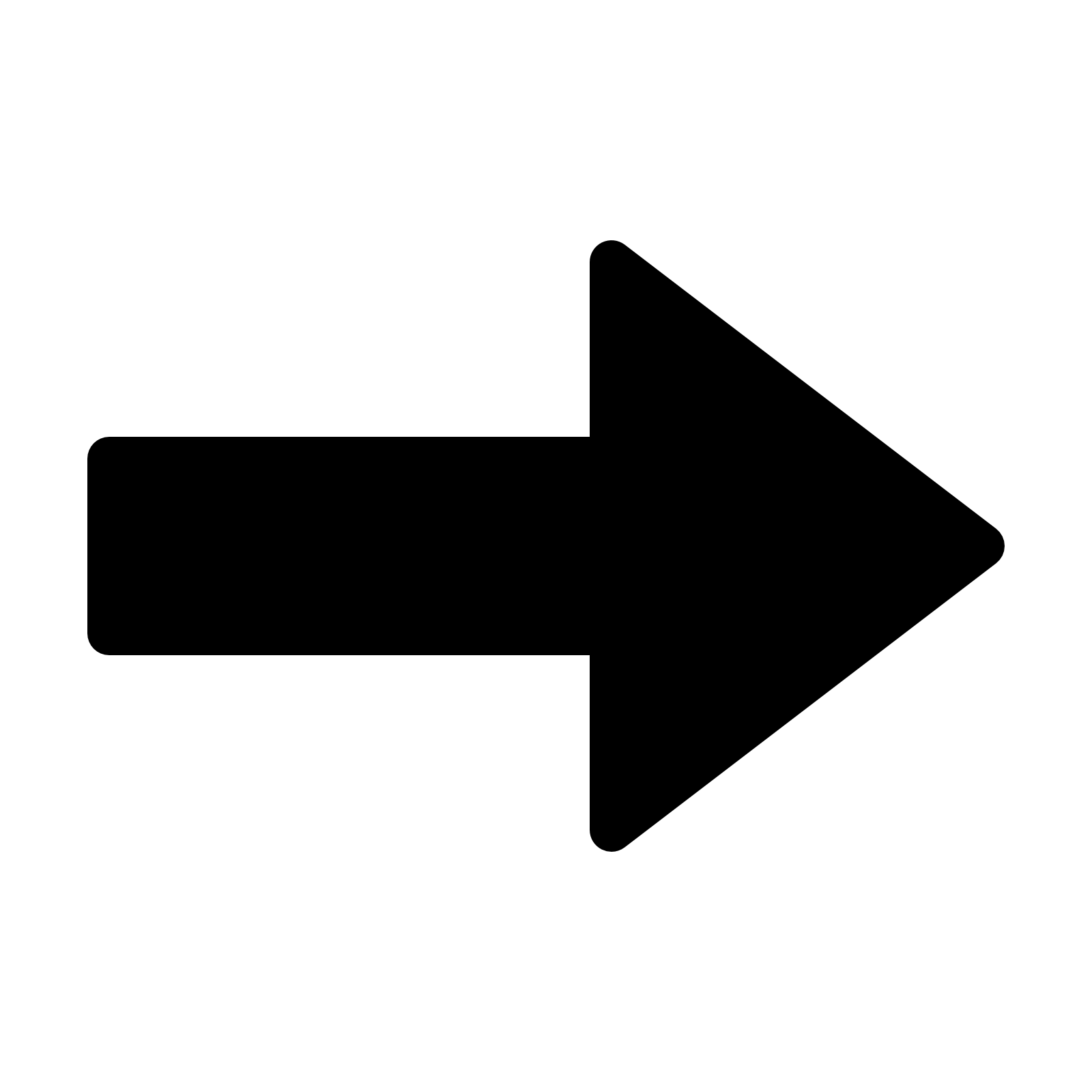}
    }
}

\begin{figure*}[t]
    \centering
    \begin{tabular}{ @{\hspace{0pt}} c @{\hspace{0pt}} c  @{\hspace{0pt}} c @{\hspace{0pt}} c @{\hspace{0pt}} c @{\hspace{0pt}} c @{\hspace{0pt}} c @{\hspace{0pt}} c @{\hspace{0pt}} c}
        \editfigbust{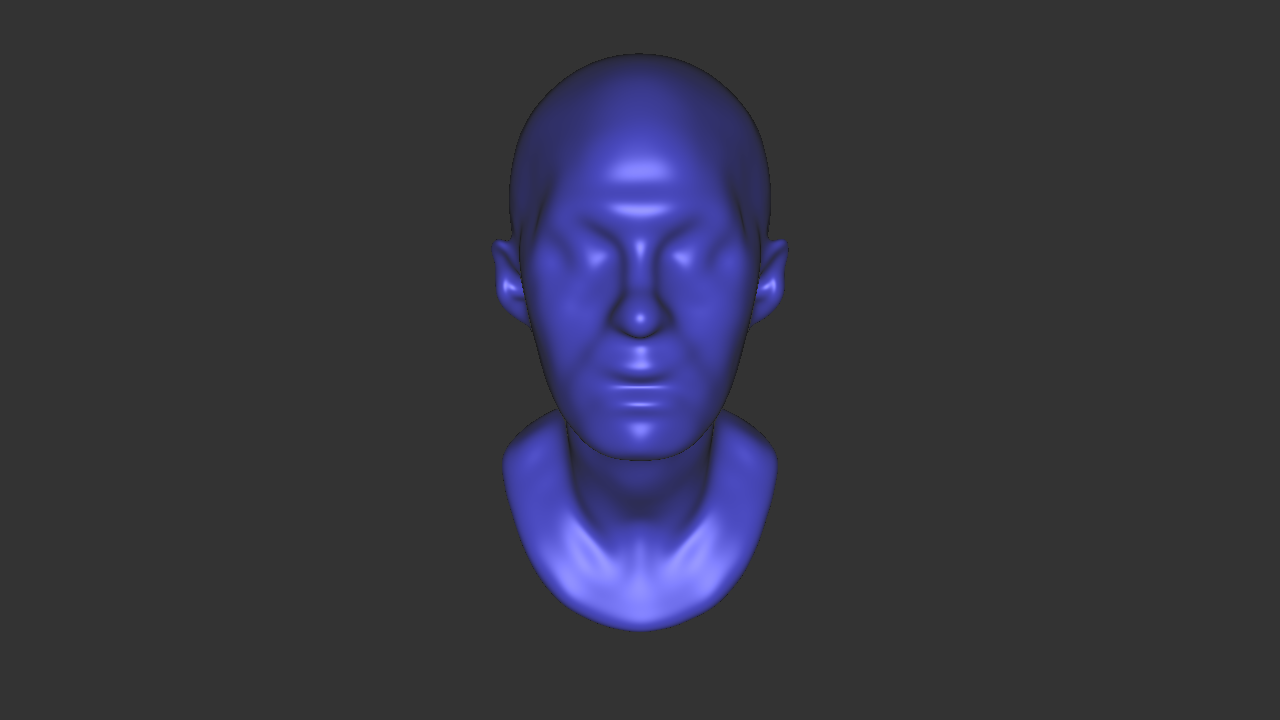} & \arrowfig & \editfigbust{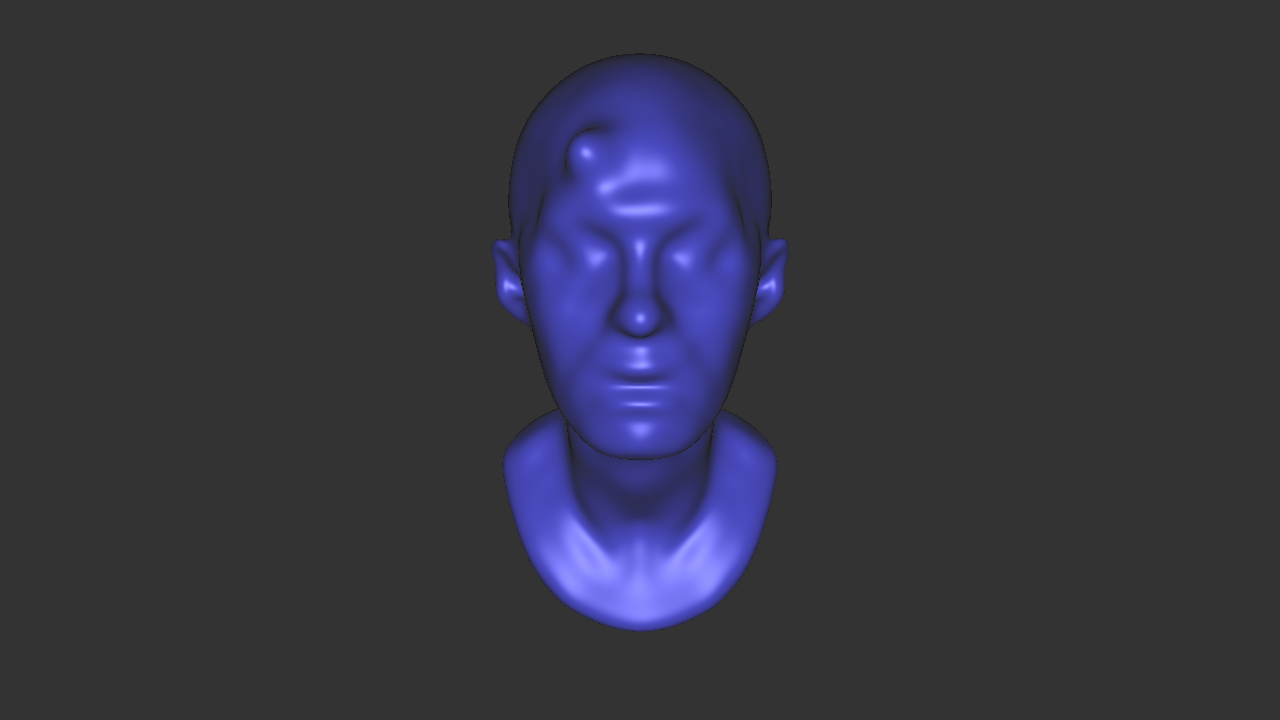} & \arrowfig & \editfigbust{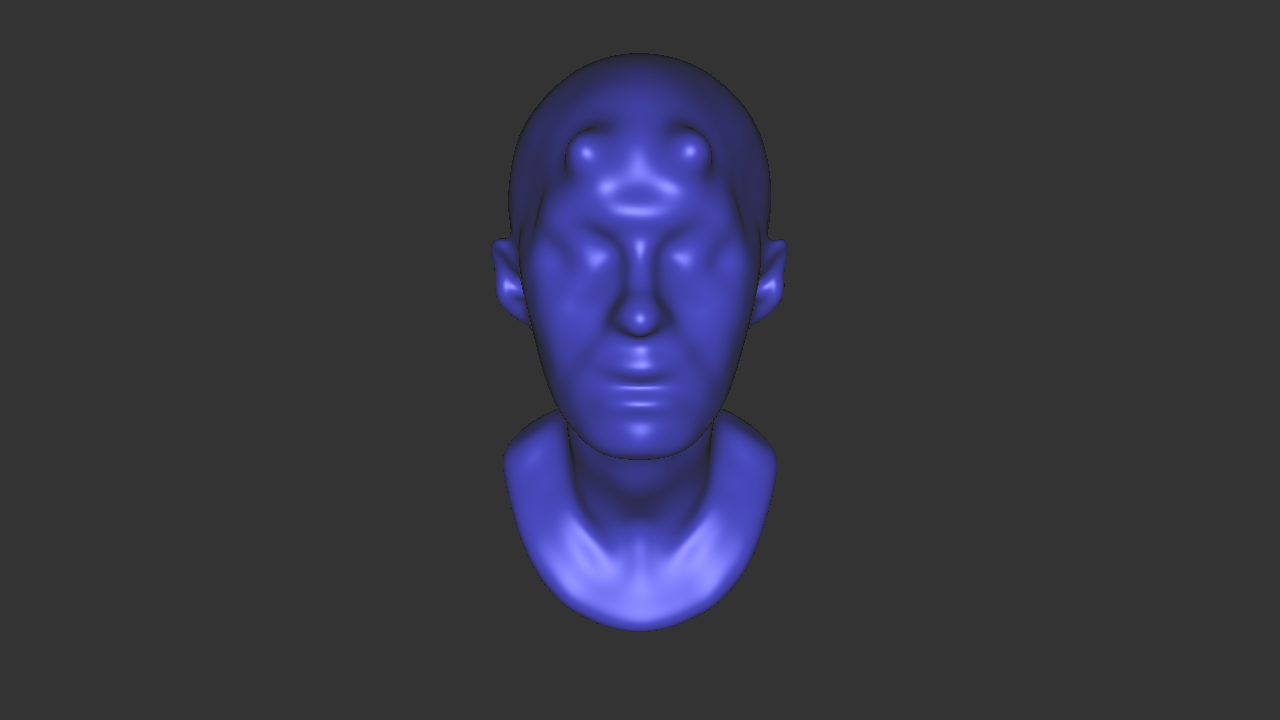} & \arrowfig & \editfigbust{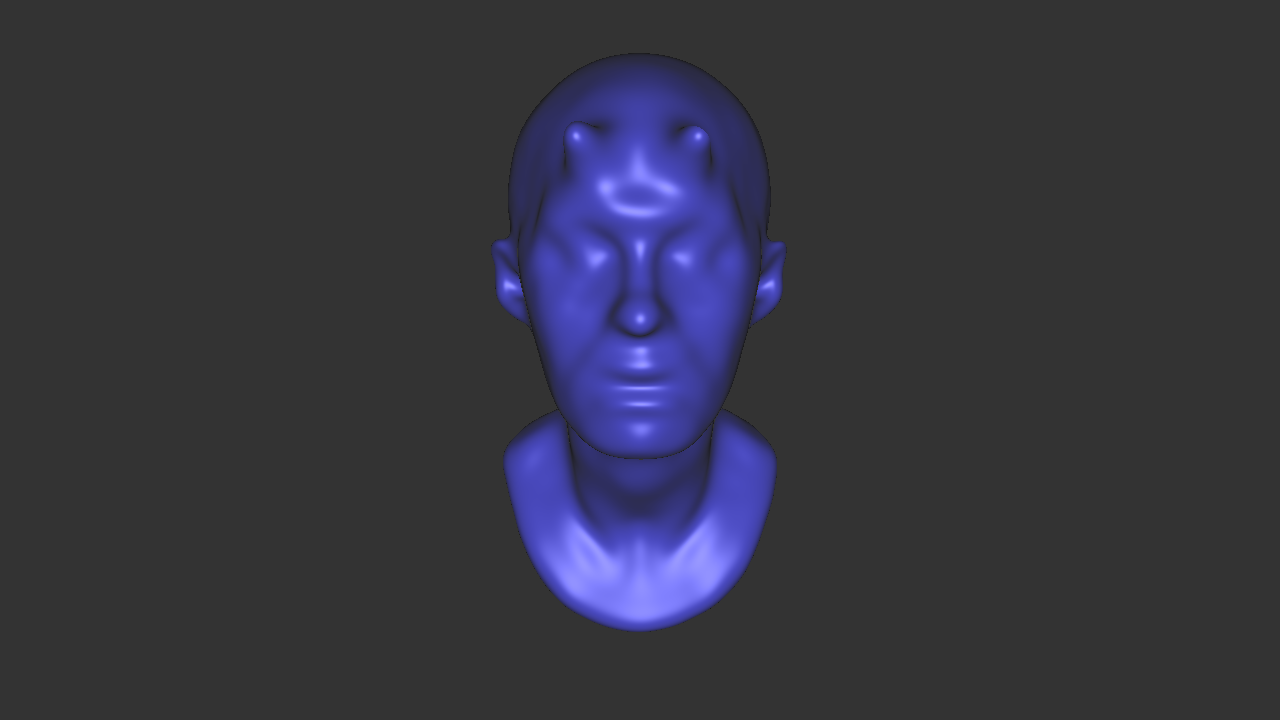} & \arrowfig & \editfigbust{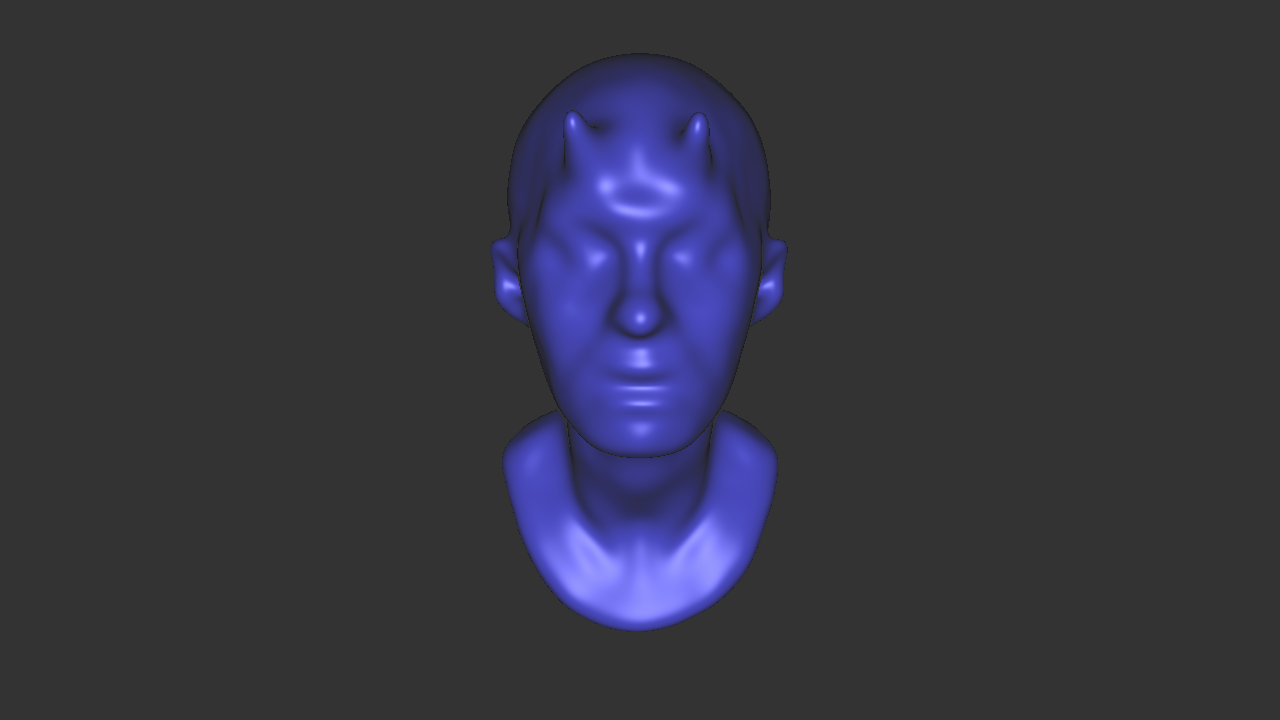} \\ \\
        \editfigbunny{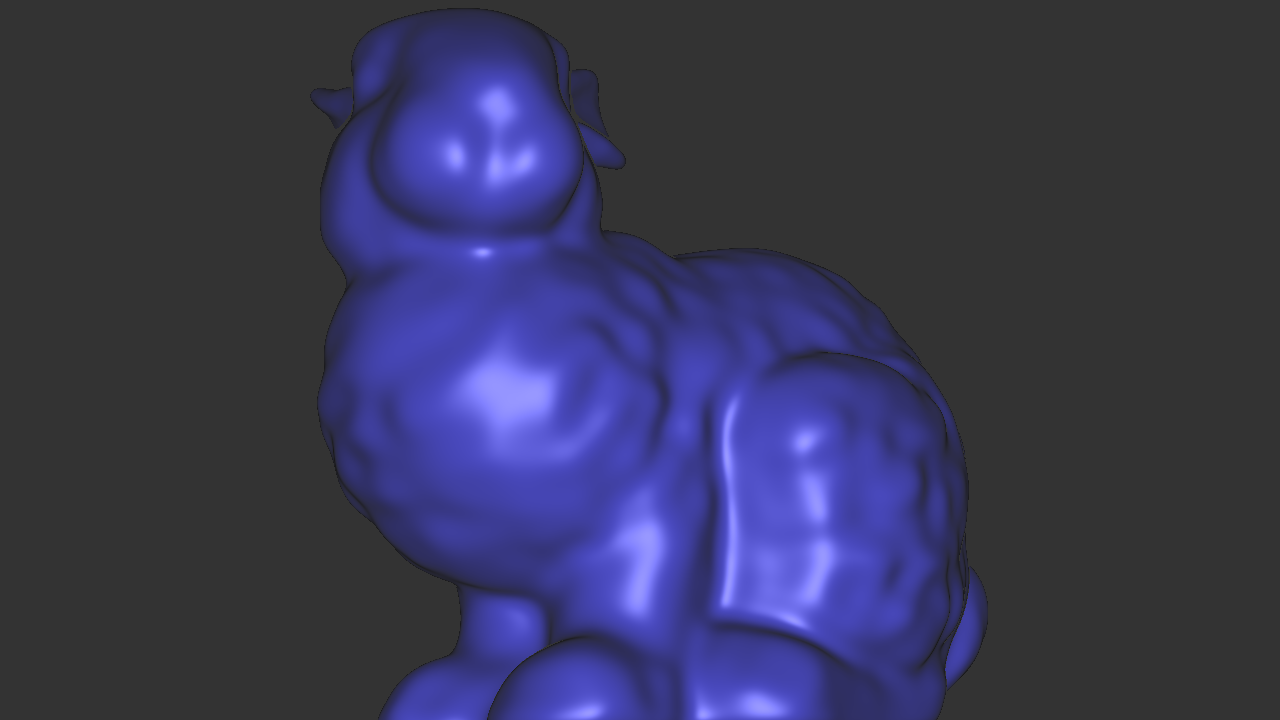} & \arrowfig & \editfigbunny{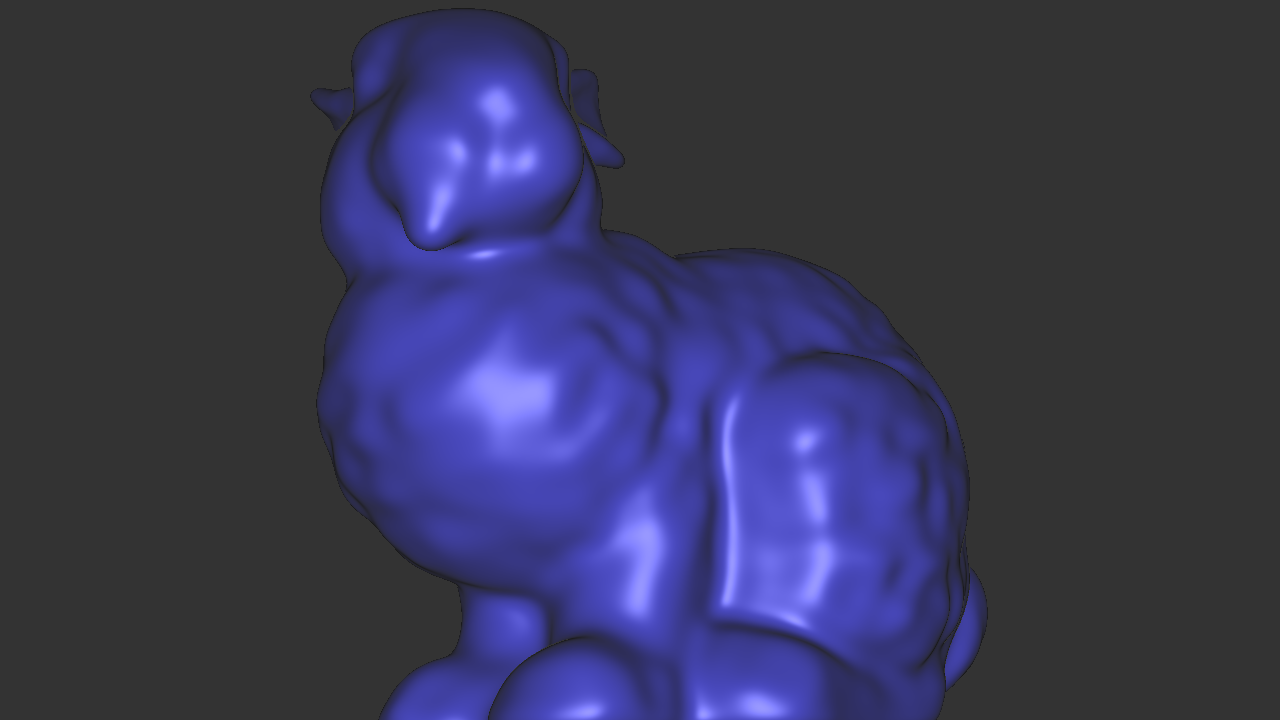} & \arrowfig & \editfigbunny{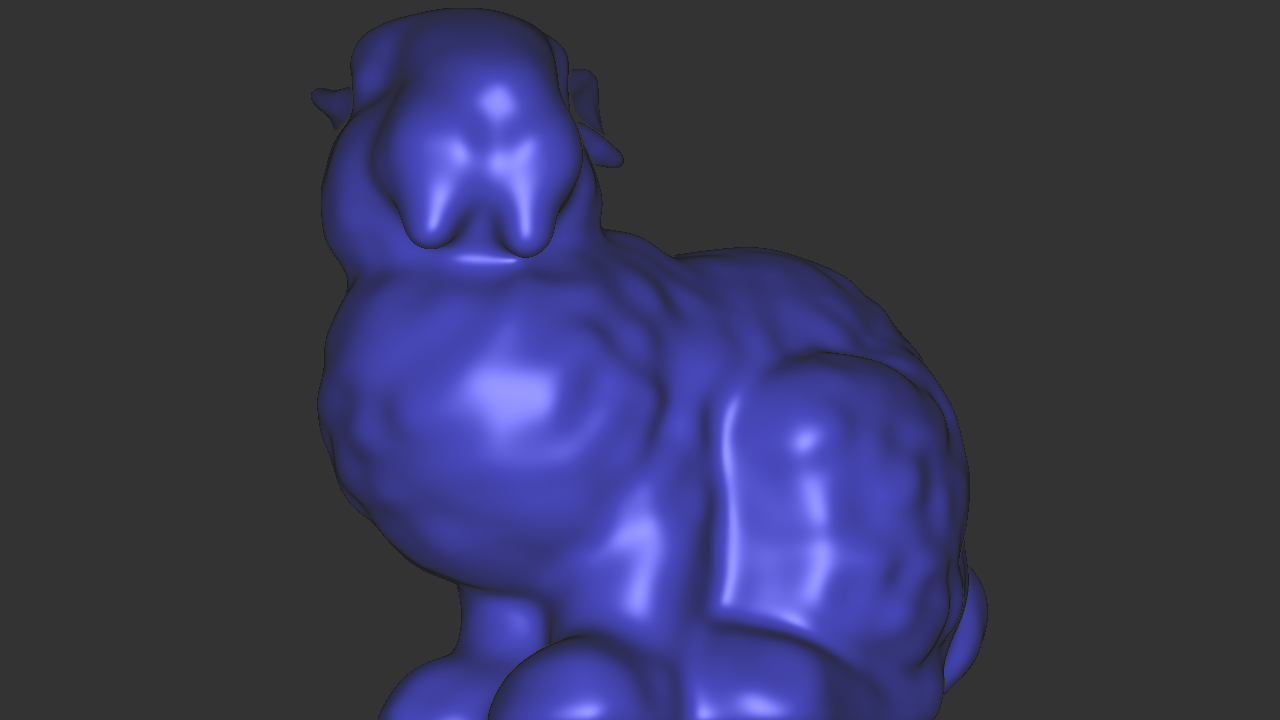} & \arrowfig & \editfigbunny{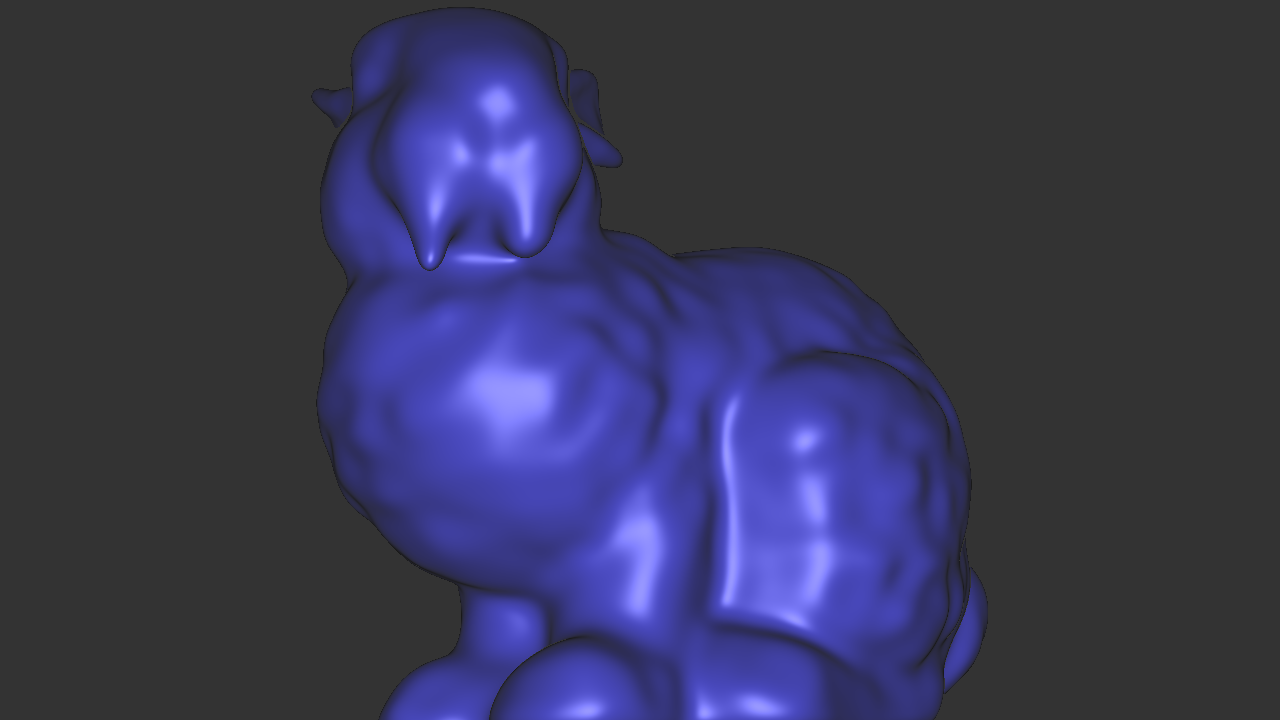} & \arrowfig & \editfigbunny{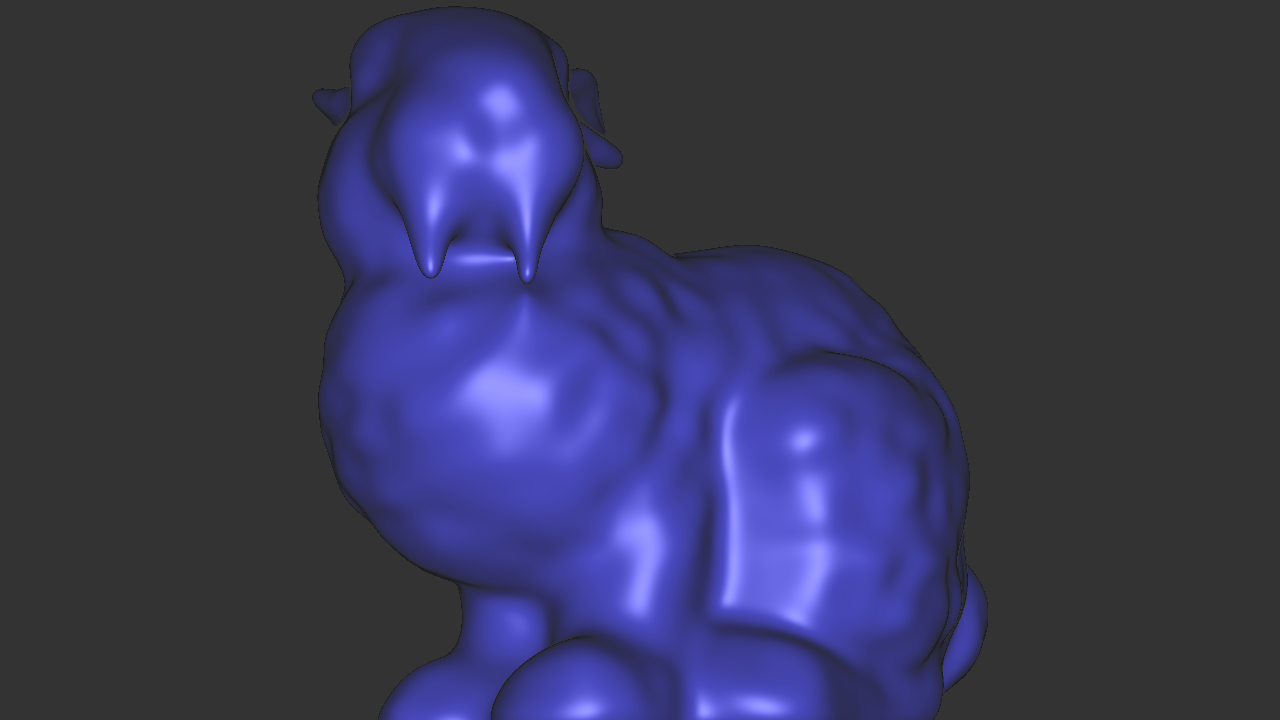} \\ \\
        \editfigpumpkin{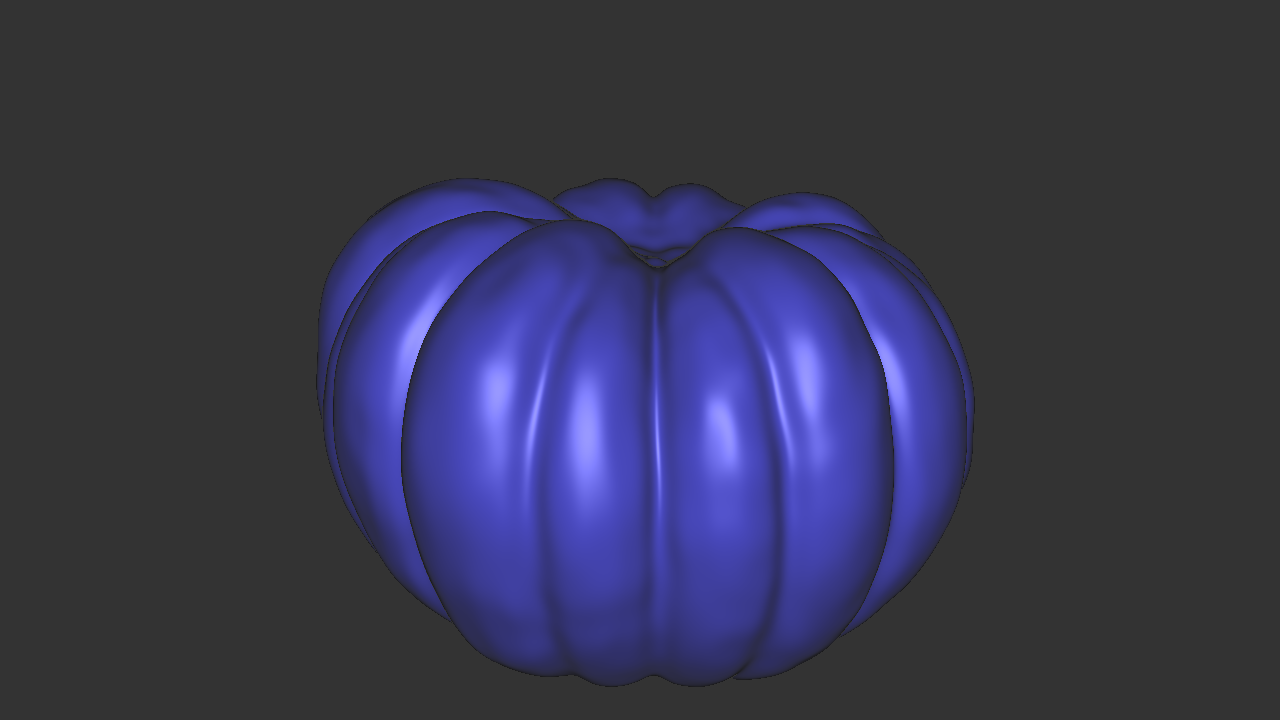} & \arrowfig & \editfigpumpkin{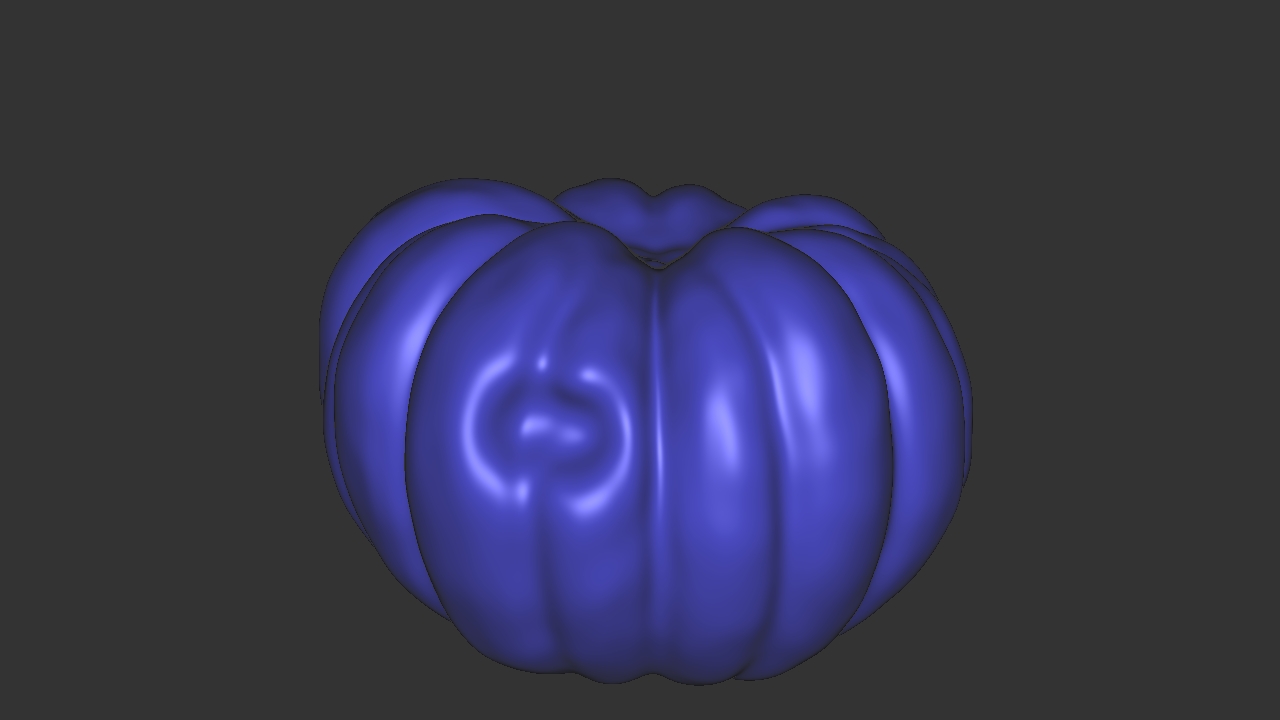} & \arrowfig & \editfigpumpkin{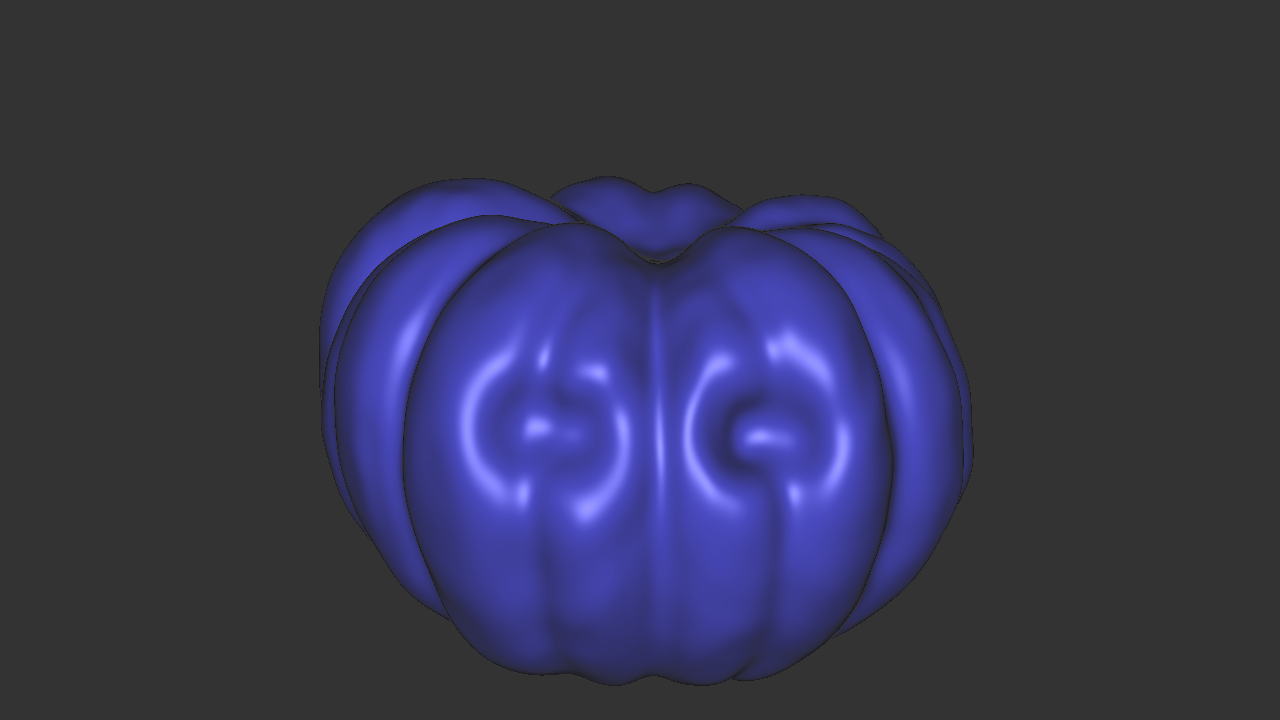} & \arrowfig & \editfigpumpkin{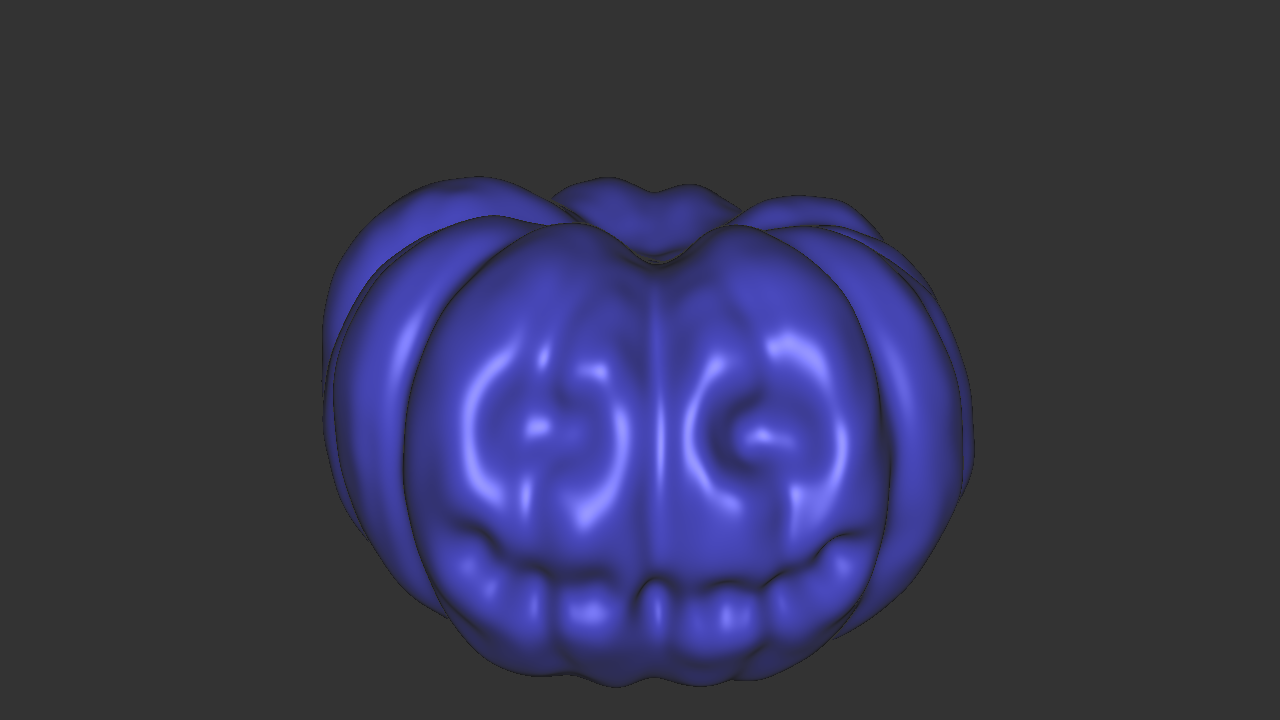} & \arrowfig & \editfigpumpkin{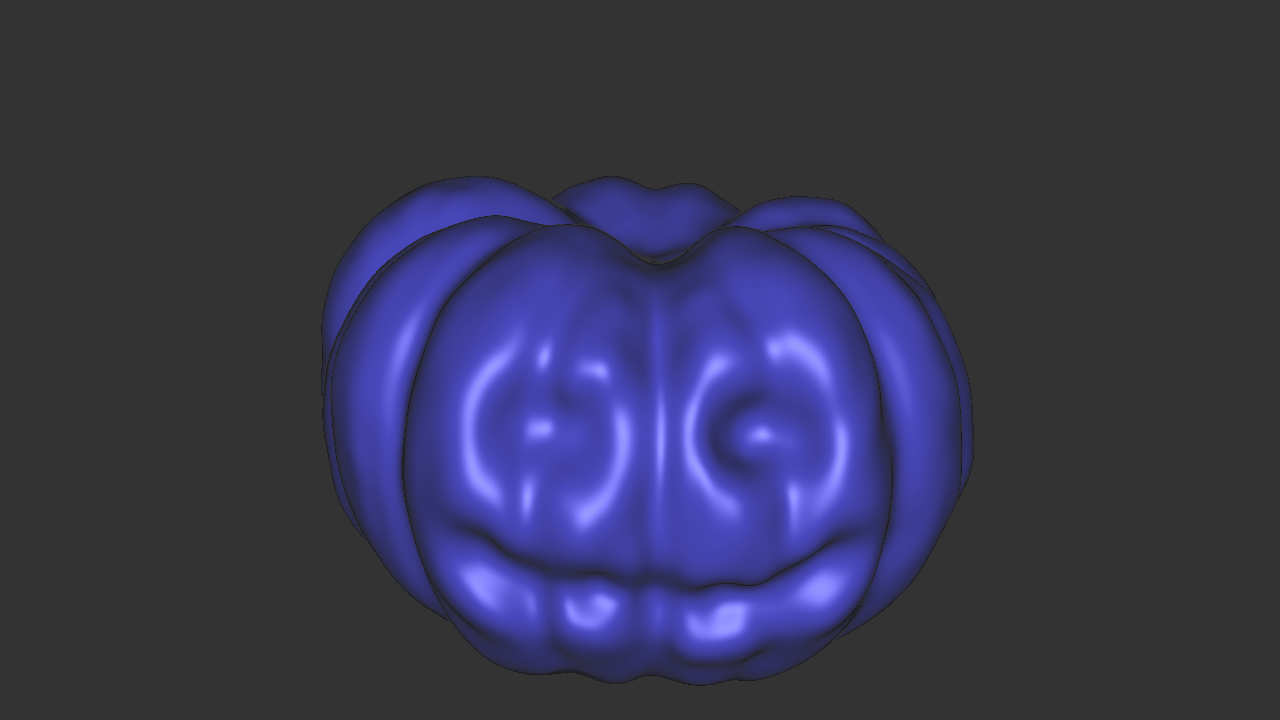}
    \end{tabular}
    \captionof{figure}{Examples of shape editing capabilities offered by our 3DNS, acting directly on the Neural SDF representation. First column: original shapes. Following columns: results of multiple edits using various brush settings, visualizing intermediate stages of the process.  All edits and renderings are done on the implicit neural representation, avoiding completely the use of triangular meshes. Images were taken from our interactive editor, which uses Sphere Tracing \cite{sphere_tracing} to render the zero-level set of the Neural SDFs. Please zoom in for details and refer to the paper's webpage (\url{https://pettza.github.io/3DNS/}) for a demo video.
}
    \label{fig:edit_fig}
\end{figure*}

\begin{abstract}
    In recent years, implicit surface representations through neural networks that encode the signed distance have gained popularity and have achieved state-of-the-art results in various tasks (\eg shape representation, shape reconstruction, and learning shape priors). However, in contrast to conventional shape representations such as polygon meshes, the implicit representations cannot be easily edited and existing works that attempt to address this problem are extremely limited. In this work, we propose the first method for efficient interactive editing of signed distance functions expressed through neural networks, allowing free-form editing. Inspired by 3D sculpting software for meshes, we use a brush-based framework that is intuitive and can in the future be used by sculptors and digital artists. In order to localize the desired surface deformations, we regulate the network by using a copy of it to sample the previously expressed surface. We introduce a novel framework for simulating sculpting-style surface edits, in conjunction with interactive surface sampling and efficient adaptation of network weights. We qualitatively and quantitatively evaluate our method in various different 3D objects and under many different edits. The reported results clearly show that our method yields high accuracy, in terms of achieving the desired edits, while at the same time preserving the geometry outside the interaction areas.
\end{abstract}

\section{Introduction}

Representing and manipulating surfaces and 3D shapes is a problem of paramount importance in many diverse applications, ranging from mechanical and architectural design to computer animation, augmented/virtual reality, and physical simulations. It thus comes as no surprise that the representations devised over the years are as many and diverse as the applications, each with their respective advantages and disadvantages. Bézier patches, B-splines and subdivision surfaces are only some of the choices, with the most ubiquitous being the polygon meshes \cite{hearn2004computer}.

Although polygon meshes offer a useful and efficient representation it is hard to model diverse topologies, as that would require the vertices or their connectivity to change. To surpass these limitations researchers have tried incorporating different geometrical representations, such as voxel grids, octrees, and implicit functions. Due to the grid (or grid-like) structure of the former two, they have been used with convolutional networks \cite{3dr2n2, hierarchical, ogn}. Nevertheless, voxel grids cannot achieve high resolution and, even though octrees address this, they, too, result in jagged models.

In the last years, given the ever-rising popularity of artificial neural networks, a new class of surface representations has been proposed, namely the Implicit Neural Representations (INRs). In this approach, the surface, which is frequently required to be closed, is represented implicitly as a level set of a neural network with one output. Several papers have presented very interesting and promising results using such representations \cite{DeepSDF, neuralfield_survey}. In most of the papers, the network tries to learn either the signed distance function or the occupancy function. Also, the network can learn only one surface or a class of surfaces by taking a class code along with the spatial coordinates as input. In contrast to 3D meshes, voxel grids, and other common representations, INRs are not coupled to spatial resolution and can be sampled at arbitrary spatial resolutions, since they are continuous functions. In this way, the memory required to accurately represent a 3D shape does not depend on the spatial resolution but on its geometric complexity. 

Despite the particularly promising results of implicit representations, there are still limitations to their usage. 
One of the most important limitations is that the shapes cannot be easily edited. This is due to the fact that in these representations the geometric structures are not represented in a local fashion. Each weight of the corresponding network affects the geometry over an unbounded region of the output space. This means that, in order to perform a localized modification on the 3D shape, generally all weights of the network need to be modified. 

This editability problem of INRs is an open challenge that has attracted very limited attention in the literature. Existing works allow for some form of interactive editing, by either optimizing the shape code fed to the network \cite{dif-net, dualsdf} or by training the networks for articulated objects \cite{articulated, a-sdf} and changing the joints' parameters (which are also fed to the network). In either case, the editing is limited inside a learned shape space and, so, these methods do not support arbitrary modifications of the shape's 3D geometry.

To overcome the aforementioned limitations, this work introduces the first method that allows interactive editing of INRs, specifically neural Signed Distance Functions (SDFs). We approach the problem from a 3D sculpting perspective, aiming at equipping INRs with functionalities that 3D modeling software have for the standard mesh representations. Our method, which we call 3D Neural Sculpting (3DNS), edits the surface modeled by the zero-level set of a neural network in a brush-based manner. As mentioned above, using a feedforward neural network to represent an SDF creates a problem of locality. For this, we propose using samples of the surface represented by the network to regulate the learning of the desired deformation. 
The source code is available at \url{https://github.com/pettza/3DNS}.

To recap, INRs are in their infancy. They have shown impressive results, but have not yet found many applications. We believe that the editability capabilities
that this paper introduces will pave the way to a plethora of applications for INRs, ranging from computer graphics, where machine learning becomes more and more popular, to robotics, where some preliminary works have given evidence about the great benefits of these representations \cite{grasping, simeonovdu2021ndf}. 

\section{Related Work} \label{sec:RelatedWork}

\subsection{Implicit Surface Representations}  

The idea to represent surfaces implicitly is by no means a new one. In fact, there have been continual attempts to use implicit representations in computer graphics and machine learning. In the shader art community, analytic implicit representations have been used to render from simple primitives to complex scenes and fractal objects \cite{quilez}. On the other hand, in the machine learning community earlier approaches have relied upon radial basis functions (RBFs) \cite{rbf_sdf} and octrees \cite{adaptive_signed_distance_fields} to express SDFs. The authors of \cite{usingparticles} use points sampled on an implicit surface to control its shape.

Recently, the use of a neural network as the function expressing the surface was proposed in by three concurrent works \cite{chen2019learning, occupancy_net, DeepSDF}, which ignited interest in these implicit neural representations. DeepSDF \cite{DeepSDF} uses a network to represent the SDF for a shape (or a shape class, using additionally a shape code as input to the network), while the other two \cite{chen2019learning, occupancy_net} express the occupancy function. Since then many more works on INRs have ensued.

The works referenced above use a regression-type loss function for training. For SDFs, this requires the computation of ground truth distances at points in space which can be difficult. Various attempts have been made to reformulate the loss function for training a neural SDF. SAL \cite{SAL} neural SDFs are trained using a loss that, nevertheless, disregards the sign of the distance, which requires careful initialization. SAL++ \cite{sal++} extends this method to utilize information about the normal vectors of the surface. The authors of \cite{controlling} incorporate samples of the level sets of the network function to the loss. Further progress was made by IGR \cite{igr}, which uses the fact that SDFs satisfy the eikonal equation to train the network as a differential equation solving network \cite{DGM}, thus requiring only uniform samples inside a bounding box of the surface and samples that lie on the surface, without computation of ground truth distances. Our loss is derived from this.

Other works have experimented with the architecture of the networks. SIREN \cite{siren}, which uses sines instead of the usual RELUs, presented promising results in a variety of tasks including surface reconstruction. Convolutional networks are used in \cite{if_net, ndf} by discretizing the input point cloud. State-of-the-art results have been attained by coupling neural networks with data structures that retain localized spatial information. Octrees, which store learnable weights are used in \cite{deepls, nglod} and a method called hash encoding is used by Instant-NGP \cite{instant-ngp}. Besides high detailed representations, NGLoD \cite{nglod} and Instant-NGP \cite{instant-ngp} achieve interactive framerates, as well.

Besides training using raw geometric data like point clouds and meshes, there have been efforts to train neural SDFs directly from images \cite{sdf_reparametrization, dvr, Oechsle2021ICCV, yariv2020multiview}. The authors of \cite{SDDF} propose an SDF variant that takes direction into account (Signed Directional Distance Function). In contrast to a neural SDF which expresses the distance approximately, they prove that their network structure ensures it expresses the SDDF of some object.

It is worth mentioning that, while we focus on neural representations of shapes, implicit representations have found success in expressing quantities other than distance functions. For example, NeRF \cite{mildenhall2020nerf} produces highly realistic images by expressing the radiance and density of a scene. A recent survey \cite{neuralfield_survey} explores these representations, to which the authors refer as Neural Fields, in depth.

\subsection{Neural SDF Editability}

The research on editability is quite limited. DualSDF \cite{dualsdf} proposes a two-level representation, one which comprises a collection of primitives (\eg spheres) and a neural SDF which share the shape space. The user is able to manipulate the fine representation of the neural SDF by specifying changes to primitives' parameters which affect the shape code. A similar process is possible with DIF \cite{dif-net}, where instead of primitives a sparse set of points is used. The authors of \cite{articulated} and those of \cite{a-sdf} both deal with articulated objects. The joints' parameters are given as input to the network and, thus, can be used to manipulate the shape.

Since the above works deform the expressed shape by proxy of the network's inputs the space of possible shapes is limited to the one learned during training. In contrast, our method allows the user to more freely change the local 3D geometry in ways that do not necessarily lie within a learned shape space,  
similar to the functionalities that until now were offered by 3D software for meshes only. 

\section{Background} \label{sec:Background}

We begin by presenting some material upon which we rely to develop our method. In Section \ref{sec:SDF_def} we give a definition of SDF, in Section \ref{sec:SIREN}, we describe SIREN \cite{siren} which is the architecture that we build upon, in Section \ref{sec:WeightNorm} we present weight normalization \cite{weight_norm} and finally in Section \ref{sec:IGR}, we describe the formulation of the adopted loss function.

\subsection{Signed Distance Functions (SDFs)} \label{sec:SDF_def}

Let $S$ be a surface in $\mathbb{R}^3$, then the (unsigned) distance of a point $\bm{x}$ to the surface is:
\begin{equation}
    \mathit{UDF}(\bm{x}, S) = \min_{\bm{y} \in S}\{ d(\bm{x}, \bm{y}) \}
    \label{eq:udf_def}
\end{equation}

\noindent where $d$ is a metric on $\mathbb{R}^3$, typically (in our case as well) the Euclidean distance.

If $S$ is closed the signed distance function is defined as follows:
\begin{equation}
    \mathit{SDF}(\bm{x}, S) =
    \begin{cases}
        \mathit{UDF}(\bm{x}, S) & \text{if } \bm{x} \text{ inside } S \\
        -\mathit{UDF}(\bm{x}, S) & \text{otherwise}
    \end{cases}
    \label{eq:sdf_def}
\end{equation}

A \textbf{Neural SDF} is a neural network that takes the spatial coordinate $\bm{x}$ as input and approximates the SDF of a surface. As a consequence, Neural SDFs use a whole trained neural network to parametrize a single shape. They effectively use a continuous function to represent a shape and in this way they are not dependent on a specific spatial resolution.

\subsection{SIREN Representation} \label{sec:SIREN}

We build upon SIREN \cite{siren}, which is a multilayer perceptron (MLP) that uses sines, instead of the usual ReLUs for its non-linearities. Evidently, sines allow the network to more accurately represent finer details. The use of sines is related to Fourier features \cite{benbarka2021seeing}, for which the aforementioned property has been theoretically explained \cite{tancik2020fourier} using the NTK framework \cite{ntk}. We chose SIREN as our architecture due to its simplicity and effectiveness. We note, however, that our method is model-independent and, therefore, could be applied to a variety of networks. The only alteration we experiment with is adding weight normalization to each layer which we discuss next.

\begin{figure}[t]
    \begin{center}
        \includegraphics[trim=100 60 100 80, clip, width=0.8\linewidth]{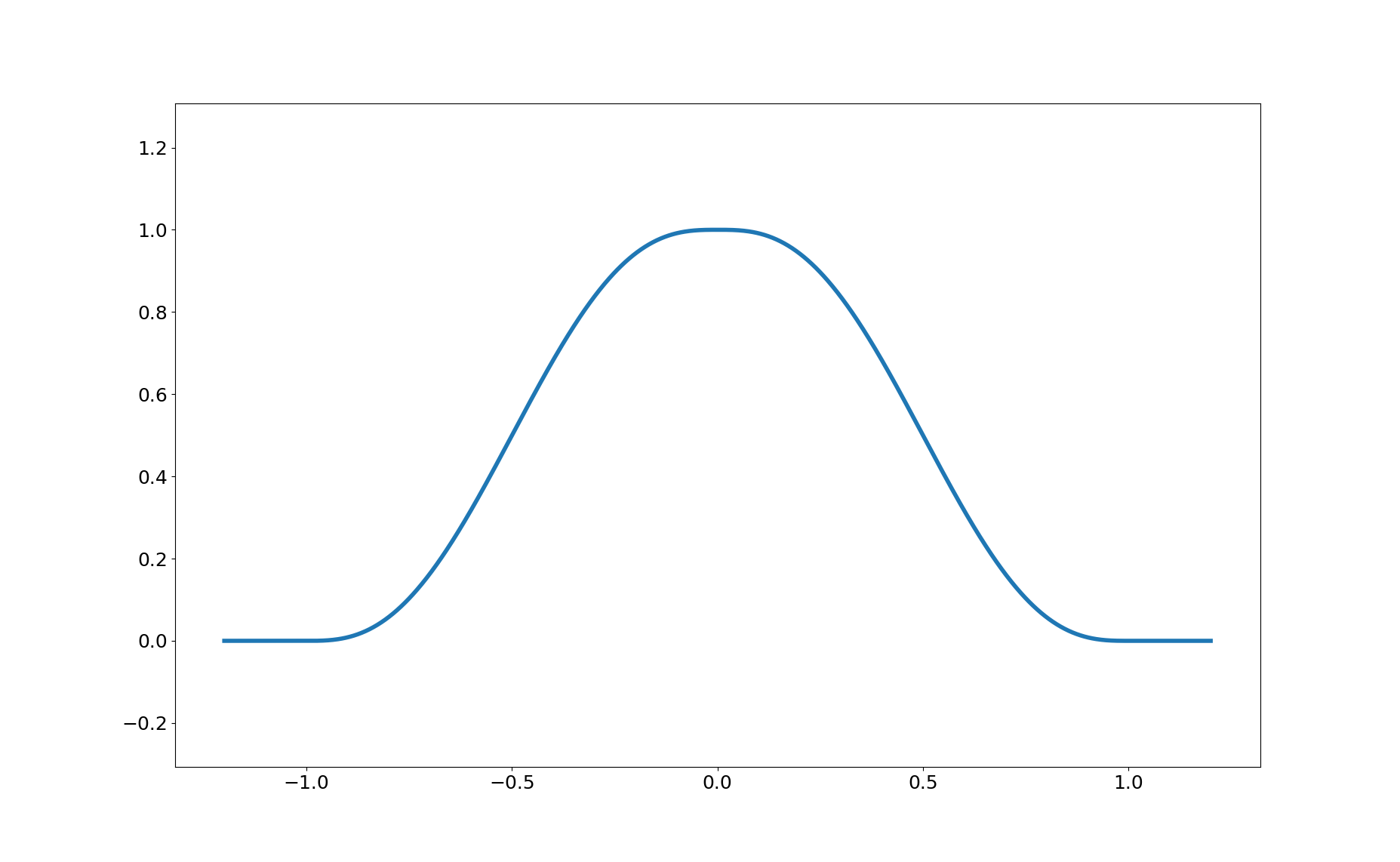}
    \end{center}
    \caption{Brush template profile. See equation \ref{eq:quintic_brush}.}
    \label{fig:quintic_brush}
\end{figure}

\subsection{Weight Normalization} \label{sec:WeightNorm}

We propose to extend SIREN by equipping it with weight Normalization, which is a memory-efficient technique proposed by \cite{weight_norm} for accelerating convergence. It works by reparametrizing the weights of a linear layer so that they are represented by a length parameter and a direction vector. This approximately
‘whitens’ the gradients during training and this makes
the 3D shape represented by the Neural SDF to update
across iterations in a more effective manner. As we show in Section \ref{sec:WeightNormAbl} applying it to SIREN has a positive effect.

\subsection{Loss Function for Neural SDF} \label{sec:IGR}

We adopt the loss function introduced by SIREN \cite{siren}  and briefly present it here for the sake of completeness. In more detail, many of the works on neural SDFs \cite{deepls, ndf, instant-ngp, DeepSDF, nglod} use a regression type loss to train the neural networks. The loss is evaluated for points sampled on the surface, near the surface, and in a bounding box around it. The ground truth signed distance needs to be computed for off-the-surface points which can be difficult, especially for non-mesh data which interest us (see Section \ref{sec:Method}). The authors of \cite{igr} propose a loss function that comprises a regression loss evaluated only on points on the surface (for which the distance is 0), a term for the normal vectors at the same points, and a term that enforces the network to have unit gradients \wrt to the input (this is commonly referred to as the eikonal term). The latter term is evaluated for points that are sampled uniformly inside a bounding box. The authors of \cite{siren}, besides some minor changes, expand the above loss with a term that penalizes small values of the neural SDF at off-surface points. In summary, the loss function $L(\theta)$ that we minimize during training is defined as:
\begin{align}
    &L(\theta) = L_S(\theta) + L_{eik}(\theta) + L_{es}(\theta) \label{eq:total_loss}, \text{ where:} \\
    &L_S(\theta) = \mathbb{E}_{p_S} \{ { \lambda_1 \, | f_{\theta}(\bm{x}) | + \lambda_2 \, g \left( \nabla_x f_{\theta}(\bm{x}), n_{\bm{x}} \right) } \} \label{eq:sdf_loss} \\
    &L_{eik}(\theta) = \lambda_3 \, \mathbb{E}_{q} \big\{ \big| \| \nabla_{\bm{x}} f_{\theta}(\bm{x}) \| - 1 \big| \big\} \label{eq:eikonal_loss} \\
    &L_{es}(\theta) = \lambda_4 \, \mathbb{E}_{q} \{ e^{- \alpha | f_{\theta}(\bm{x}) | } \} \label{eq:empty_space_loss}
\end{align}

\noindent where $\lambda_1, \lambda_2$, $\lambda_3$ and $\lambda_4$ are balancing weights (set to $1.5 \cdot 10^3$, $5$, $2.5$, $5$ respectively), $S$ is the target surface, $p_S$ is a distribution on the surface, $q$ is uniform distribution in a bounding box, $\theta$ is the parameter vector, $f_{\theta}$ is the network function, $g$ is the cosine distance, $n_{\bm{x}}$ is the normal vector at $\bm{x}$ and $\alpha$ a large positive number (set to $100$). 
$L_S$ encompasses the regression and normal terms, $L_{eik}$ is the eikonal term, and $L_{es}$ is the term described last.

\section{Proposed 3D Neural Sculpting} \label{sec:Method}

Our goal is to deform a surface (represented as a neural SDF) around a selected point on it in a manner similar to what is possible for meshes with 3D sculpting software. We refer to the selected point as the \textit{interaction point}, to the area around it as the \textit{interaction area} and to the process in general as an \textit{interaction} or \textit{edit}. The main problem in our case is how to enforce locality. Generally, a change to the parameters of a neural network is expected to affect its output for an unbounded region of the input space. Consequently, naively trying to train the network only where the surface needs to change will ostensibly distort the rest of the surface as well, which is what we confirm through experimentation in Section \ref{sec:MeshComp}. In order to ameliorate this adverse effect as much as possible, we produce the point cloud that is to be used to evaluate the loss of equation \ref{eq:sdf_loss} by including both samples from the surface that the network already represents (we call these model samples) and samples from the desired deformation. After discarding samples from the former set that are close to the interaction point we use the union with the latter set for training. In Section \ref{sec:SurfaceSampling} we describe our surface sampling, then we present our formulation of sculpting brushes in Section \ref{sec:Brushes} and then, in Section \ref{sec:InteractionSampling} we describe how we sample the interaction area.

\begin{figure*}[t]
    \centering
    \begin{subfigure}{0.14\linewidth}
        \includegraphics[height=0.9in]{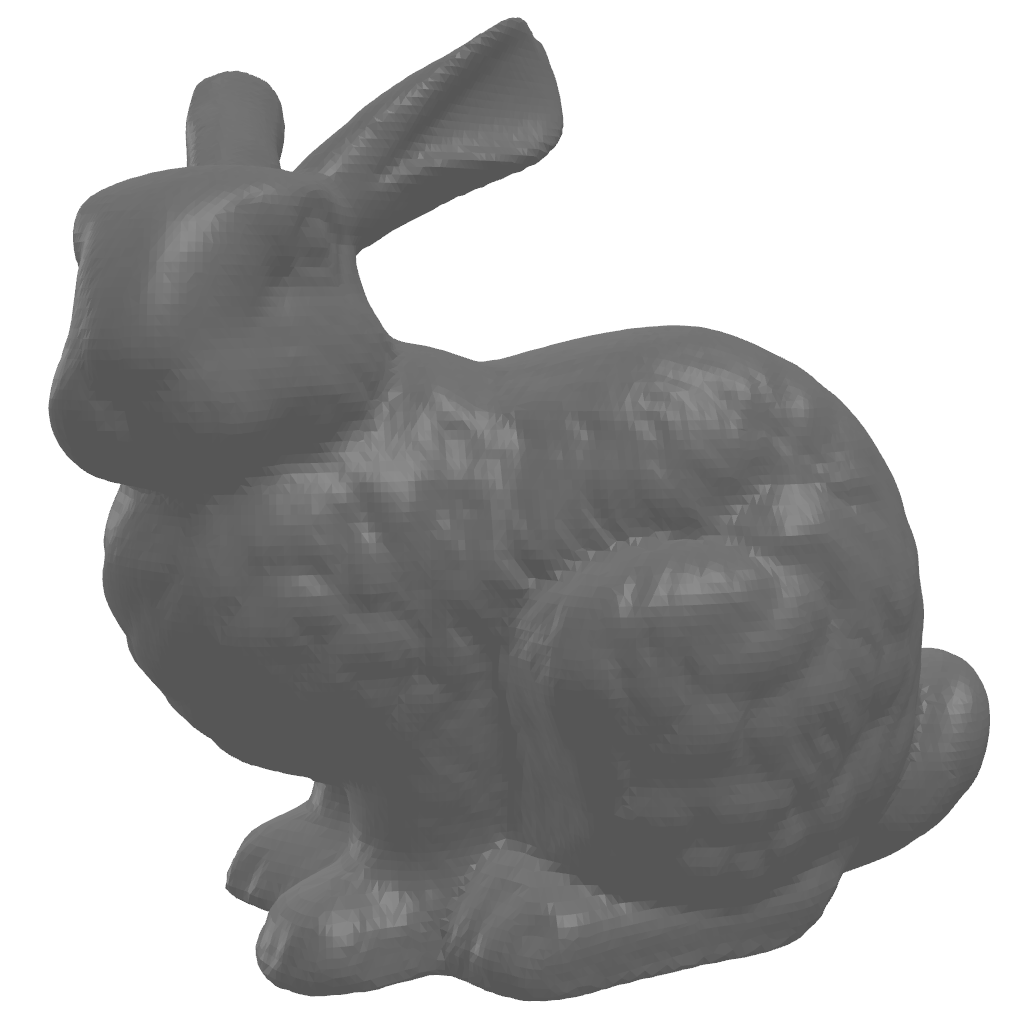}
        \caption{Bunny}
        \label{fig:bunny_shape}
    \end{subfigure}
    \begin{subfigure}{0.17\linewidth}
        \includegraphics[height=0.9in]{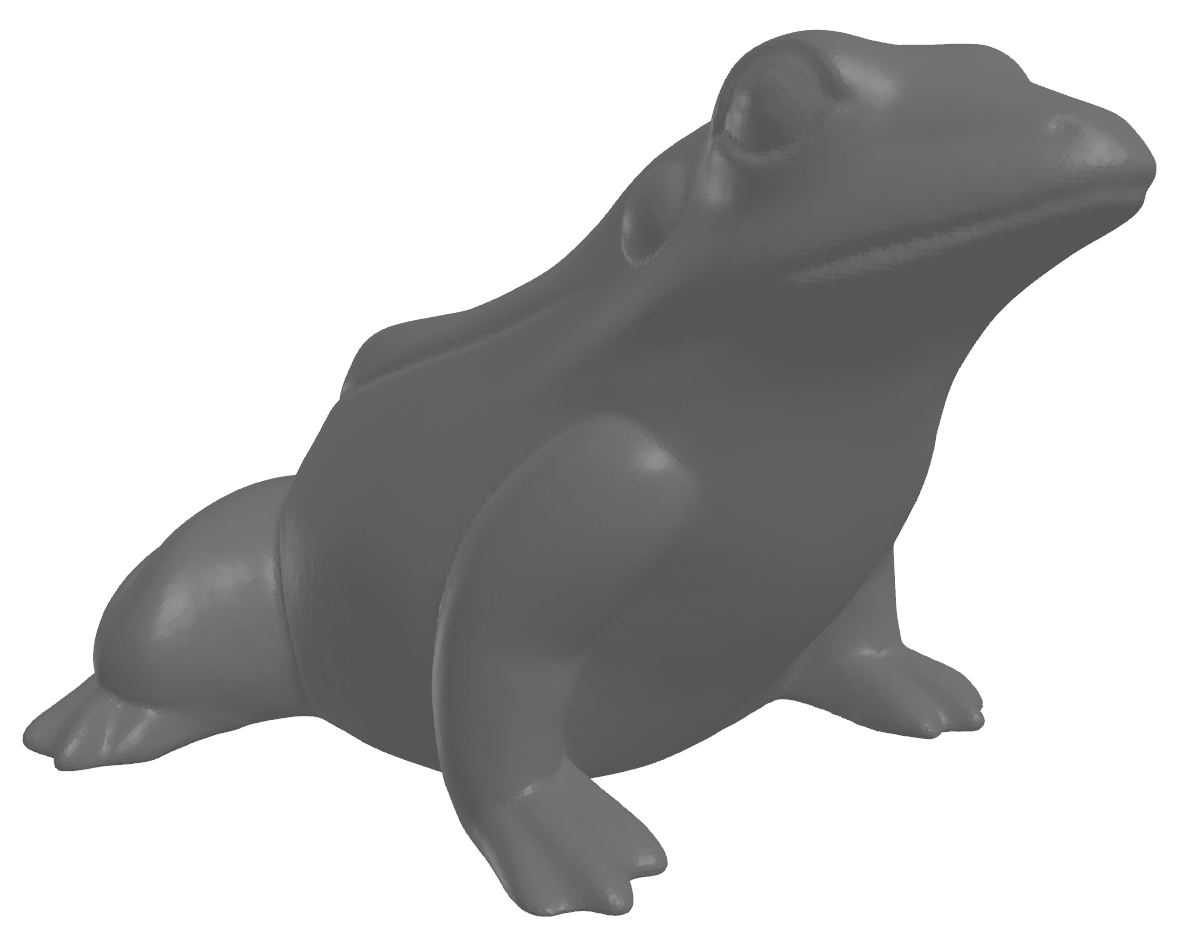}
        \caption{Frog}
        \label{fig:frog_shape}
    \end{subfigure}
    \begin{subfigure}{0.076\linewidth}
        \includegraphics[height=0.9in]{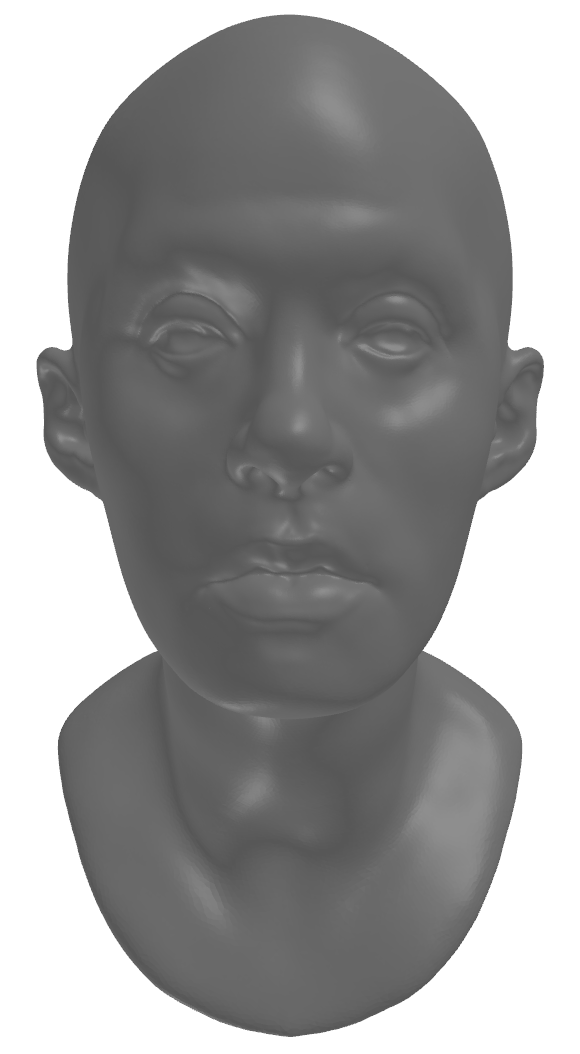}
        \caption{Bust}
        \label{fig:bust_shape}
    \end{subfigure}
    \begin{subfigure}{0.158\linewidth}
        \includegraphics[height=0.9in]{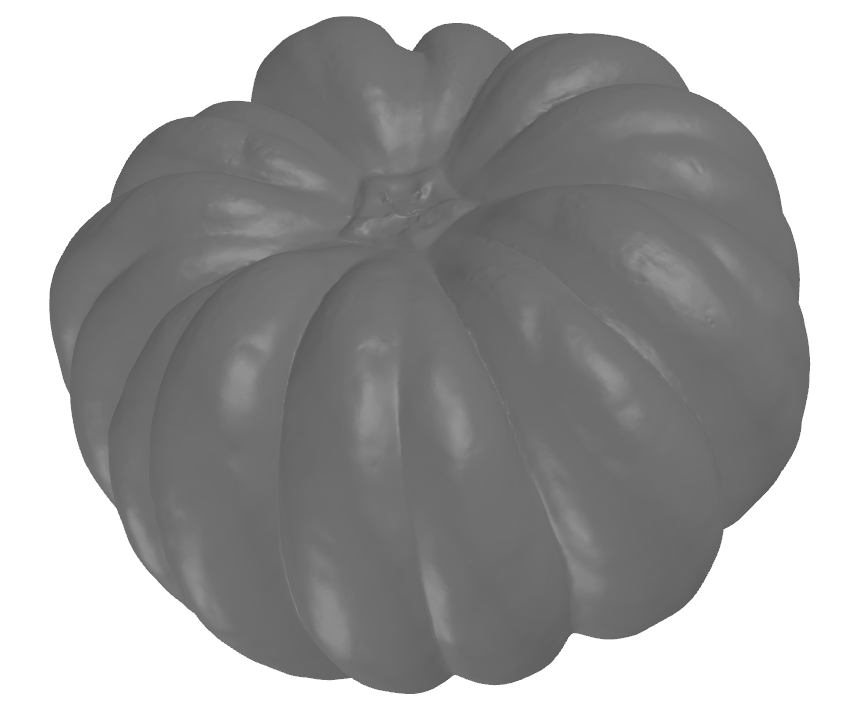}
        \caption{Pumpkin}
        \label{fig:pumpkin_shape}
    \end{subfigure}
    \begin{subfigure}{0.153\linewidth}
        \centering
        \includegraphics[height=0.9in]{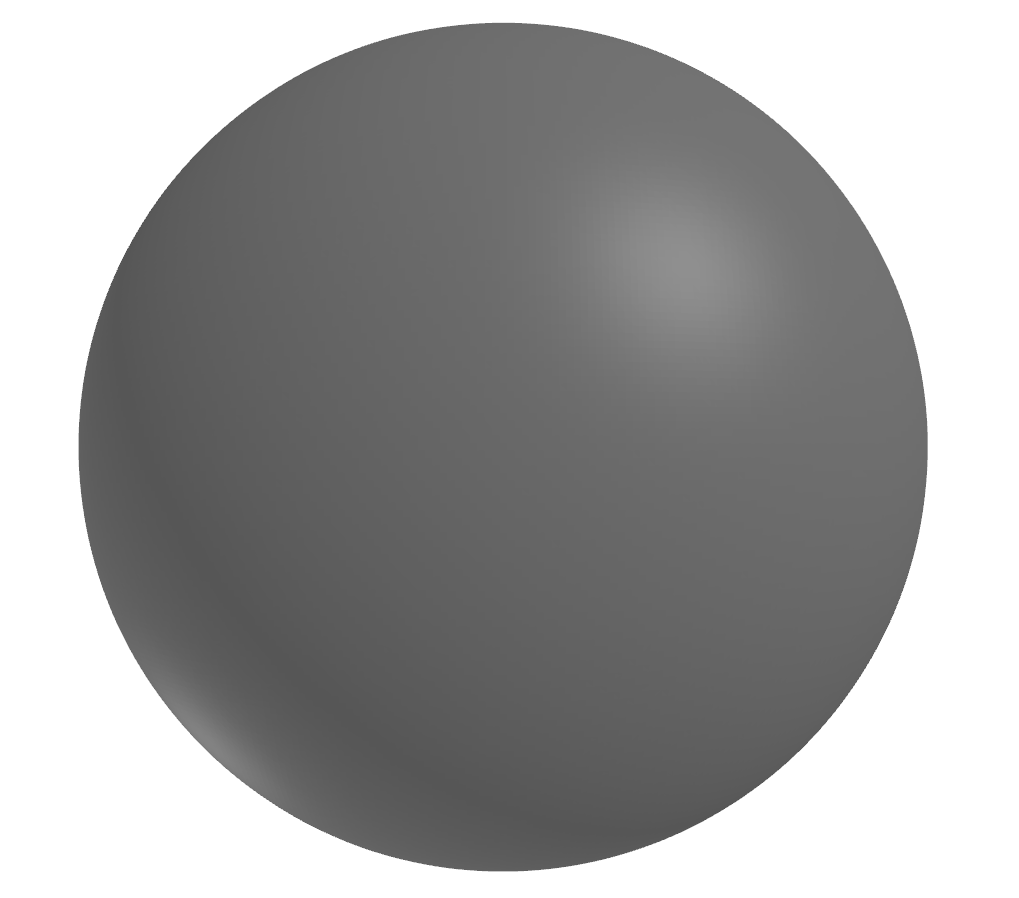}
        \caption{Sphere}
        \label{fig:sphere_shape}
    \end{subfigure}
    \begin{subfigure}{0.153\linewidth}
        \includegraphics[height=0.9in]{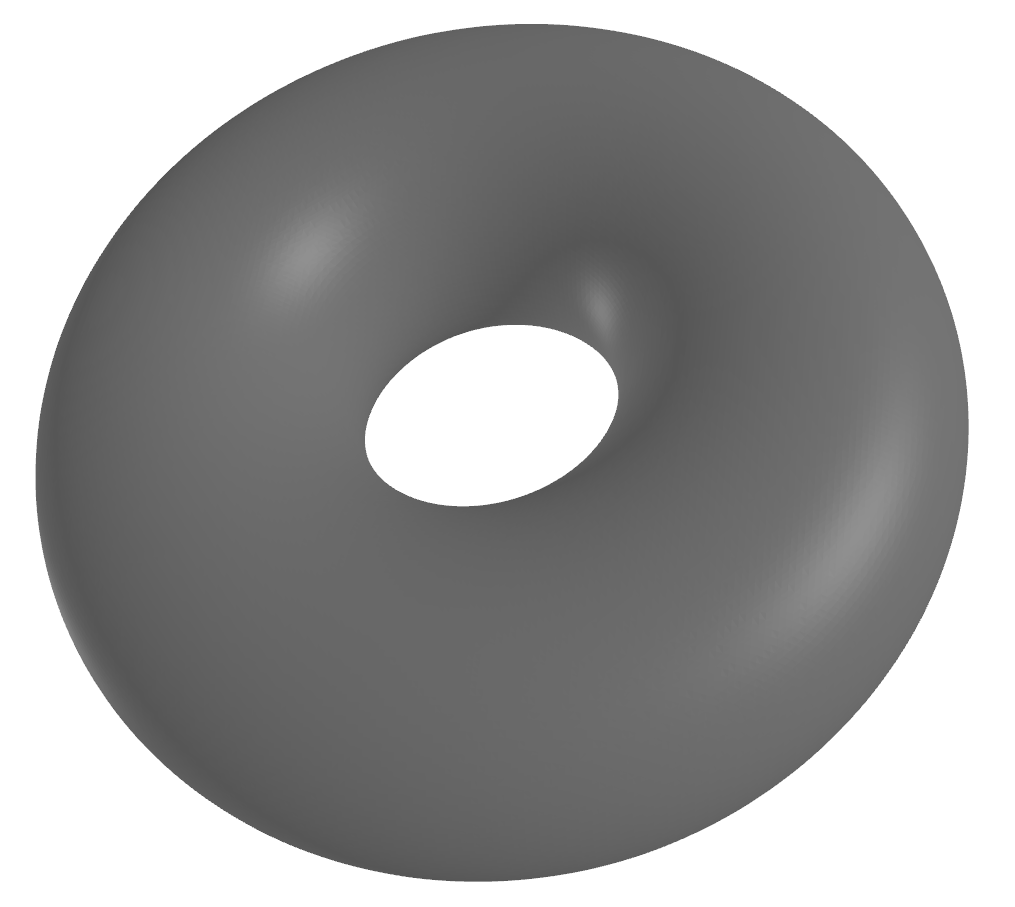}
        \caption{Torus}
        \label{fig:torus_shape}
    \end{subfigure}
    \caption{The dataset of 3D shapes that we use in our experiments. Please zoom in for details.}
    \label{fig:meshes}
\end{figure*}

\subsection{Surface Sampling} \label{sec:SurfaceSampling}

 Two prior works \cite{controlling, ndf} have proposed similar algorithms that sample the zero-level set of a neural network function. They work by sampling points uniformly inside a bounding box and then projecting them on the level set with generalized Newton-Raphson iterations. For a true SDF, this would move a point to the surface point closest to it.

A naive way to produce samples for each training iteration would be to use this algorithm. However, this approach has two major drawbacks. Firstly, it is time inefficient. Secondly, the distribution of the resulting samples can be quite non-uniform. Inspired by \cite{ndf} where the sample set is extended by perturbing existing samples with gaussian noise we produce the samples for the next iterations using the ones we already have. To that purpose, we add to each point a vector sampled uniformly from the tangent disk and then reproject them on the surface using the aforementioned procedure. We opt for the tangential disks, instead of gaussians, so that we explore the surface as much as possible without moving too far from it. The radius of the tangent disks is a hyperparameter we set to 0.04.

\begin{figure}[b]
    \centering
    \includegraphics[trim=70 40 90 20, clip, width=0.7\linewidth]{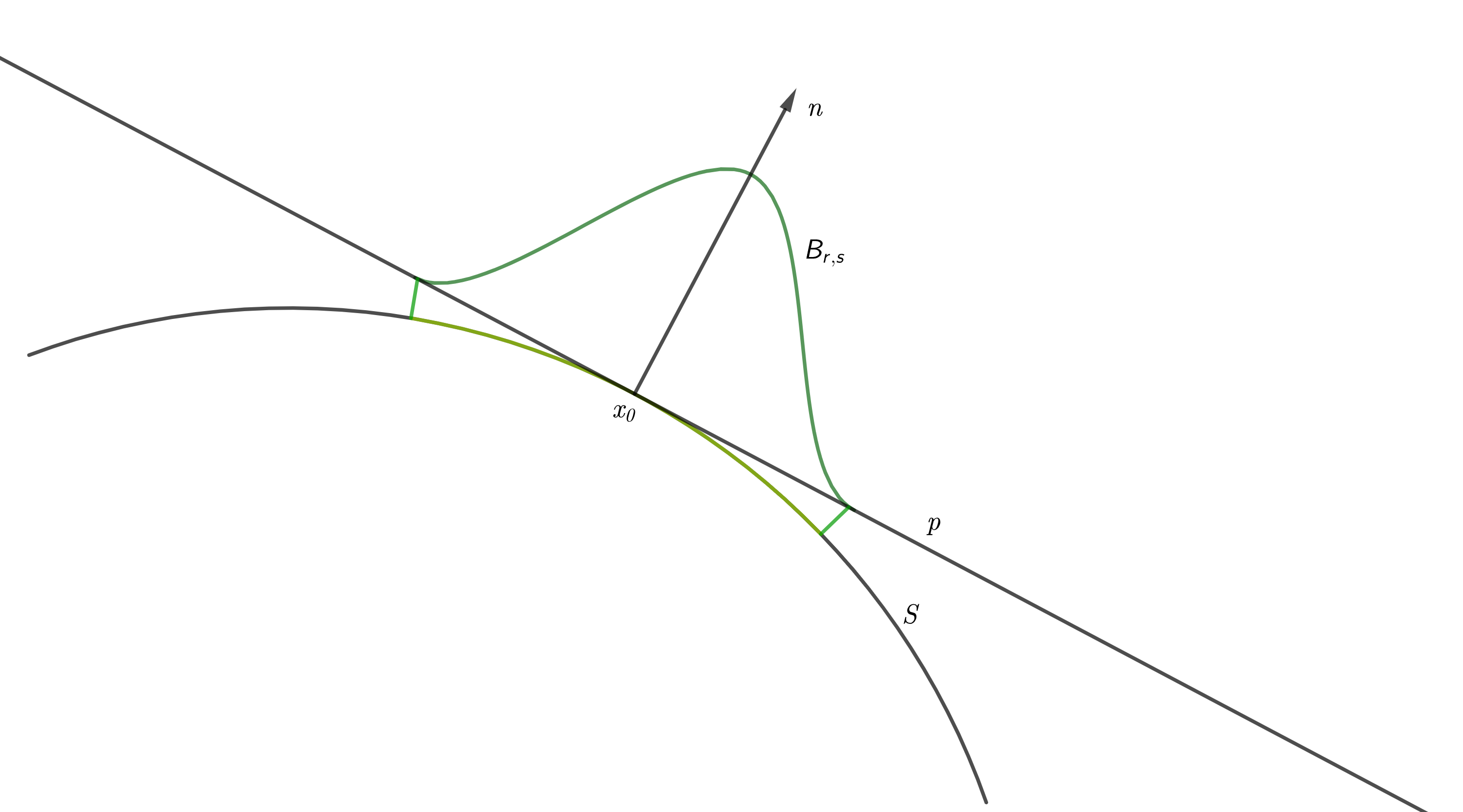}
    \caption{The brush application is demonstrated above. $S$ is the surface, $\bm{x}_0$ is the interaction point, $\bm{n}$ is the normal vector at $\bm{x}_0$, $p$ is the tangent plane and $B_{r, s}$ is the brush function whose graph over $p$ is the dark green curve.}
    \label{fig:interaction_sampling}
\end{figure}

This way of sampling forms a Markov Chain \cite{sheldon_prob}. Naturally, the stationary pdf distribution is of interest. It is relatively easy to see that the requirements of Theorem 1 of \cite{diaconis1997markov} 
(regarding the existence of the stationary distribution of a continuous space Markov Chain)  
are satisfied and, hence, the stationary distribution exists and has support over the surface. It is hard to theoretically reason about the shape of the distribution. However, we provide experimental results in Section \ref{sec:PDFEstimation}, which demonstrate that our sampling process produces more uniformly distributed samples. Uniformness guarantees that every surface region is included equally during training.

\newlength{\brushcolwidth}
\setlength{\brushcolwidth}{0.65in}

\newlength{\brusfigwidth}
\setlength{\brusfigwidth}{\dimexpr2\tabcolsep+\brushcolwidth\relax}

\newcommand{\brushfig}[2]{%
    \parbox[c]{\brusfigwidth}{%
        \includegraphics[width=\linewidth]{Figures/brush_parameters/r#1_i#2.png}%
    }%
}

\begin{figure*}[t]
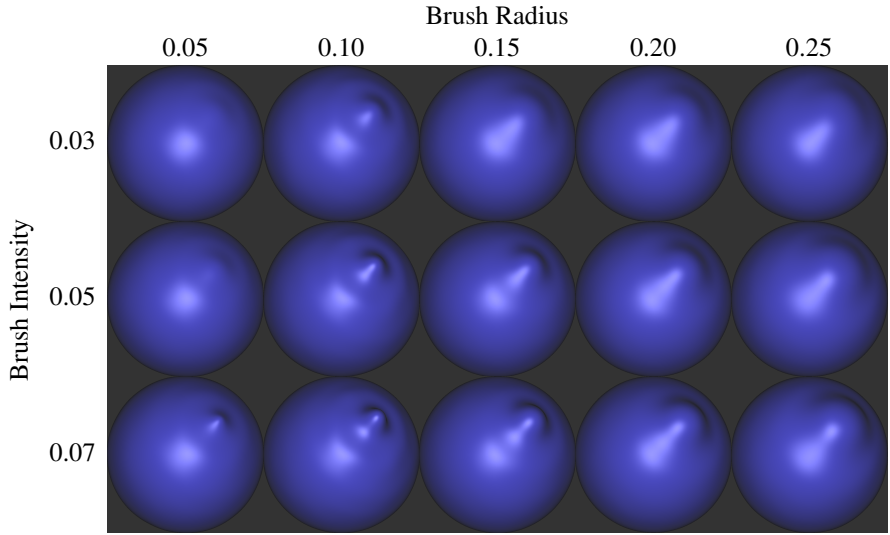

    \centering
    \begin{NiceTabular}
        {w{c}{1pt} w{c}{14pt} *{5}{w{c}{\brushcolwidth}}}
        & & \Block{1-5}{Brush Radius}\\
        & & 0.05 & 0.10 & 0.15 & 0.20 & 0.25 \\
        \Block{3-1}{\rotate Brush Intensity} & \Block{1-1}{0.03} & \brushfig{0.05}{0.03} & \brushfig{0.10}{0.03} & \brushfig{0.20}{0.03} & \brushfig{0.20}{0.03} & \brushfig{0.25}{0.03} \\
        & \Block{1-1}{0.05} & \brushfig{0.05}{0.05} & \brushfig{0.10}{0.05} & \brushfig{0.15}{0.05} & \brushfig{0.20}{0.05} & \brushfig{0.25}{0.05} \\
        & \Block{1-1}{0.07} & \brushfig{0.05}{0.07} & \brushfig{0.10}{0.07} & \brushfig{0.15}{0.07} & \brushfig{0.20}{0.07} & \brushfig{0.25}{0.07} \\
    \end{NiceTabular}
    \caption{Example of the effect of a supported bumping brush (causing an outward local deformation) on the same interaction point on a sphere using different values for the radius and intensity. A similar effect is supported for a denting brush (causing an inward local deformation).}
    \label{fig:brush_params}
\end{figure*}

\subsection{Brushes} \label{sec:Brushes}

We define a brush template as a $C^1$ (or higher) positive 2D function defined over the unit disk centered at the origin which reaches a maximum value of 1 and vanishes at the unit circle (ideally its gradient and its higher derivatives vanish as well). Suppose $b_T(\bm{x})$ is a brush template, then the properties above are summarized as follows:
\begin{align}
    &b_T : \{ \bm{x} \in \mathbb{R}^2 \, | \, \| \bm{x} \| \leq 1 \} \rightarrow \mathbb{R} \\
    &\max\{ b_T(\bm{x}) \} = 1 \\
    &\| \bm{x} \| = 1 \Rightarrow b_T(\bm{x}) = 0 \left( \wedge \: \nabla_x b_T(\bm{x}) = \bm{0} \right) 
\end{align}

We can, then, define a brush family $B_{r, s}$ parametrized over radius $r$ and intensity $s$:
\begin{align}
    B_{r,s}(\bm{x}) = s \, b_T \left( \frac{\bm{x}}{r} \right), \, r \in \mathbb{R^+},  s \in \mathbb{R}
\end{align}

We show the behaviour of different radii and intensities in Section \ref{sec:BrushParams}. Notice that a positive value for the intensity creates a bump on the surface, while a negative value creates a dent. In our experiments, we use the following brush template whose profile is shown in Figure \ref{fig:quintic_brush}. For other types, please refer to the supplementary material.
\begin{align}
    &b_T(\bm{x})=
    \begin{cases}
        P(1 - \| \bm{x} \|) & \text{if } \| \bm{x} \| < 1 \\
        0 & otherwise
    \end{cases} \label{eq:quintic_brush} \\
    &P(x) = 6 x^5 - 15 x^4 + 10 x^3 
\end{align}

In order to apply the brush at a point on the surface, we consider it defined on the tangent plane at that point. Due to the implicit function theorem \cite{vector_calc} we can express the zero-level set of the network function, in a region of that point, as the graph of a 2D function over the same plane. We can, thus, apply the brush by simply adding the brush function to the latter. As we will see, even though this is not important for computing the samples' positions, we use it to compute the correct deformed normals.

\begin{figure*}[t]
    \centering
    \begin{minipage}[c]{.22\linewidth}
        \centering
        \includegraphics[width=0.7\textwidth]{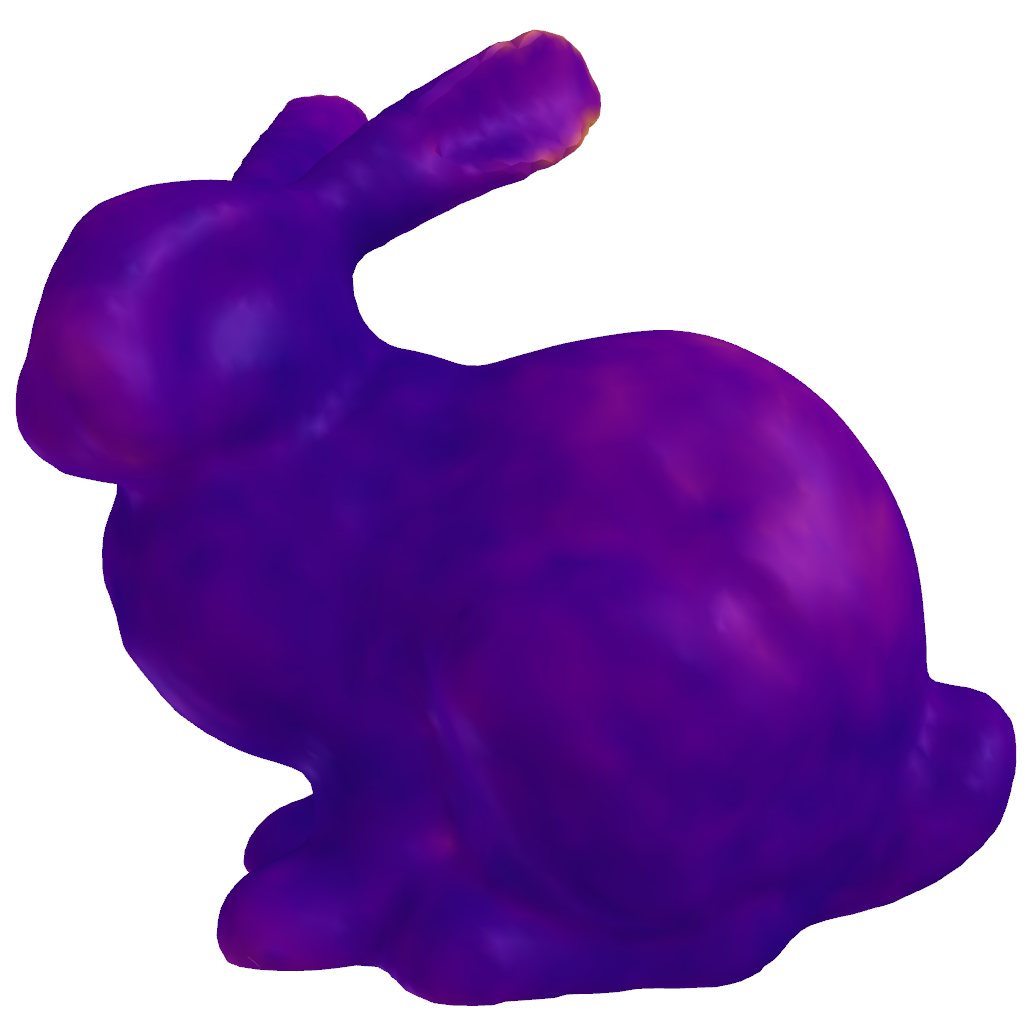}
        \par
        \includegraphics[width=\textwidth]{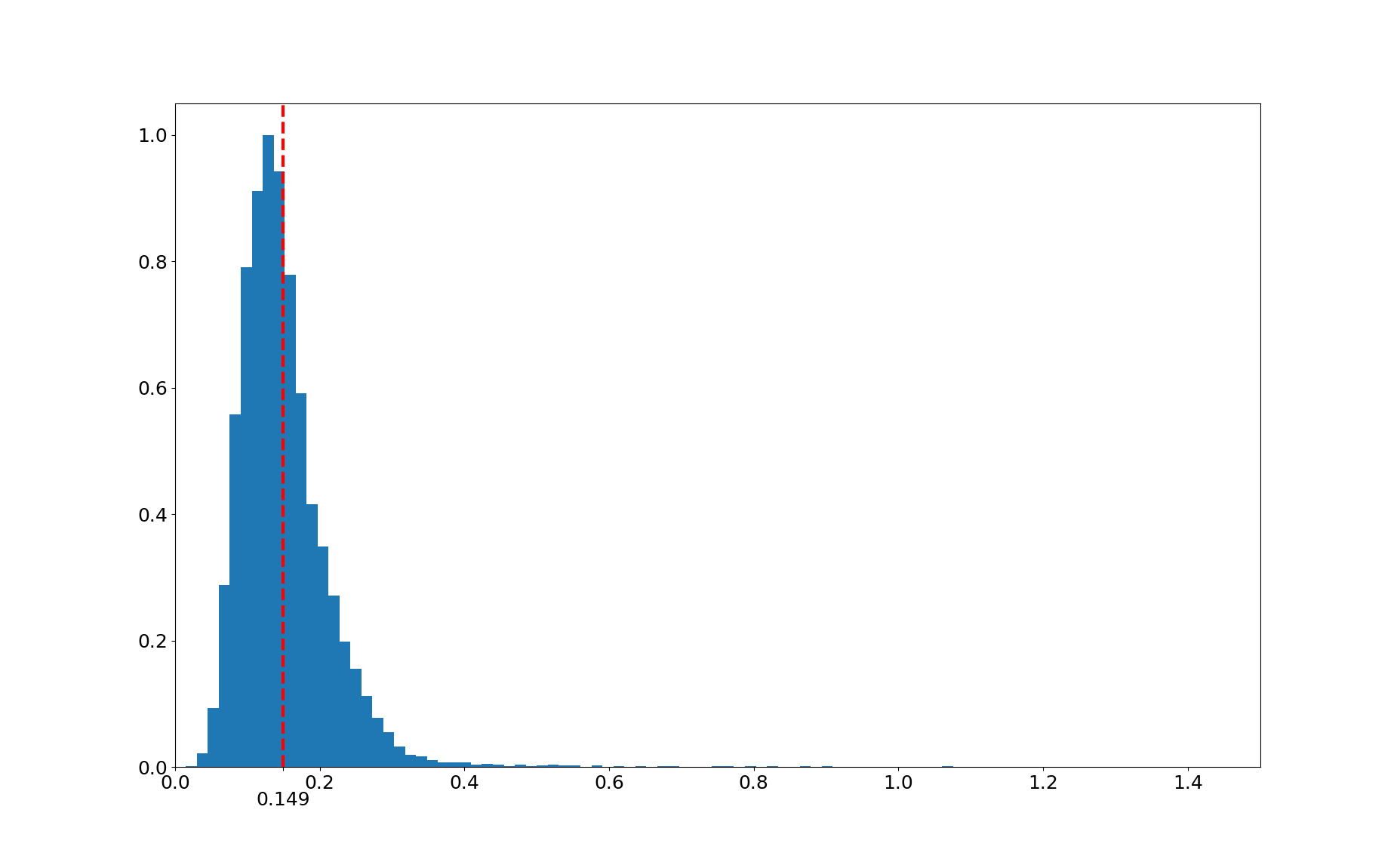}
        \label{fig:bunny_pdf_ours}
    \end{minipage}
    \begin{minipage}[c]{.22\linewidth}
        \centering
        \includegraphics[width=0.7\textwidth]{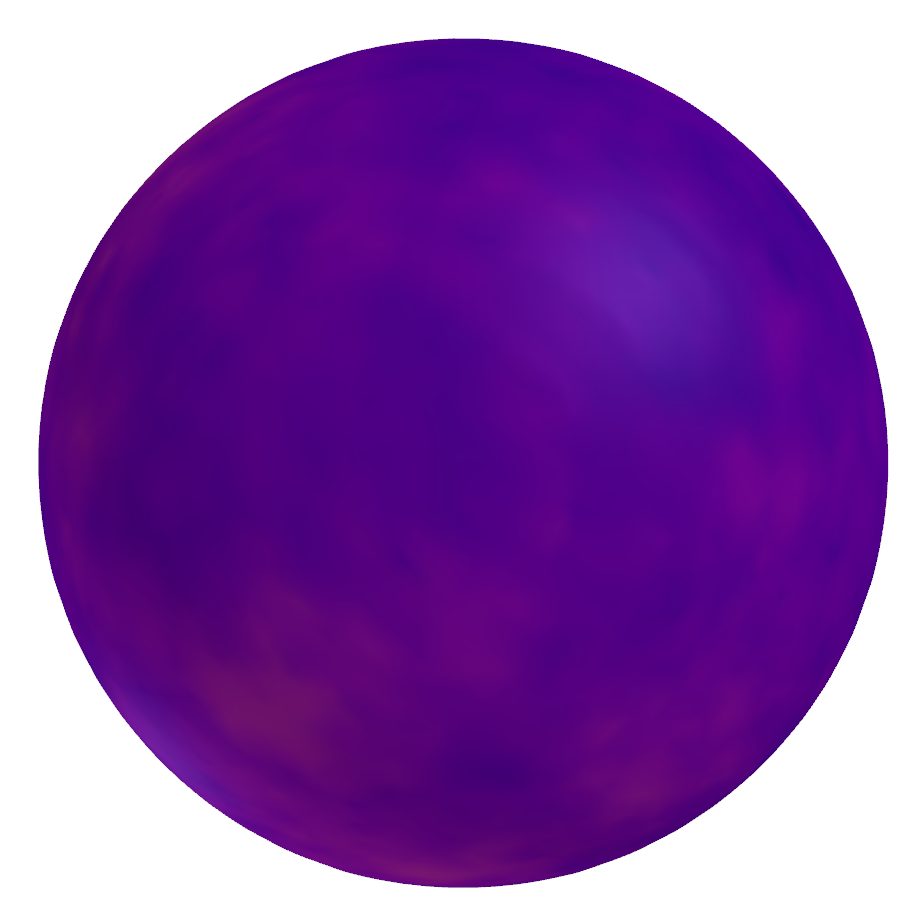}
        \par
        \includegraphics[width=\textwidth]{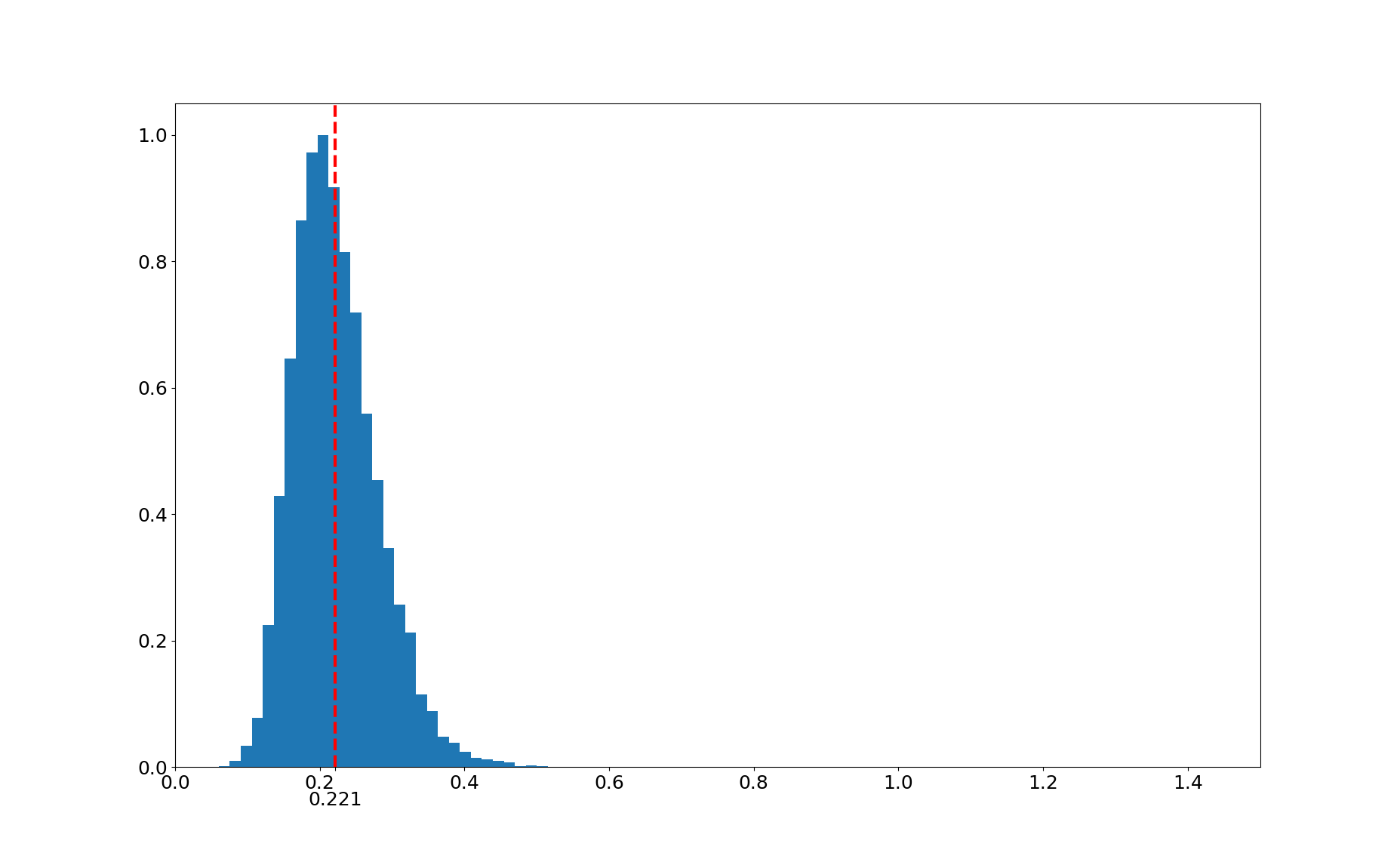}
        \label{fig:sphere_pdf_ours}
    \end{minipage}
    \begin{minipage}[c]{.22\linewidth}
        \centering
        \includegraphics[width=0.7\textwidth]{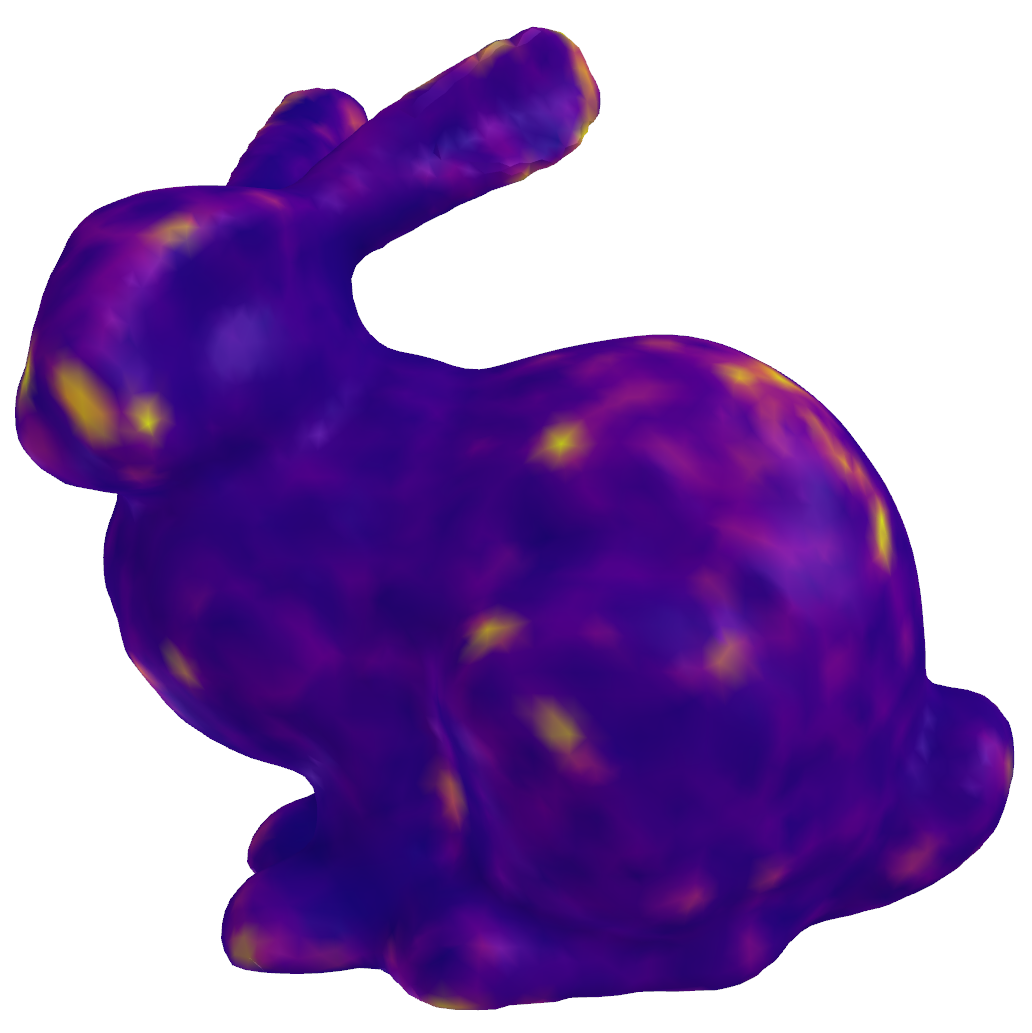}
        \par
        \includegraphics[width=\textwidth]{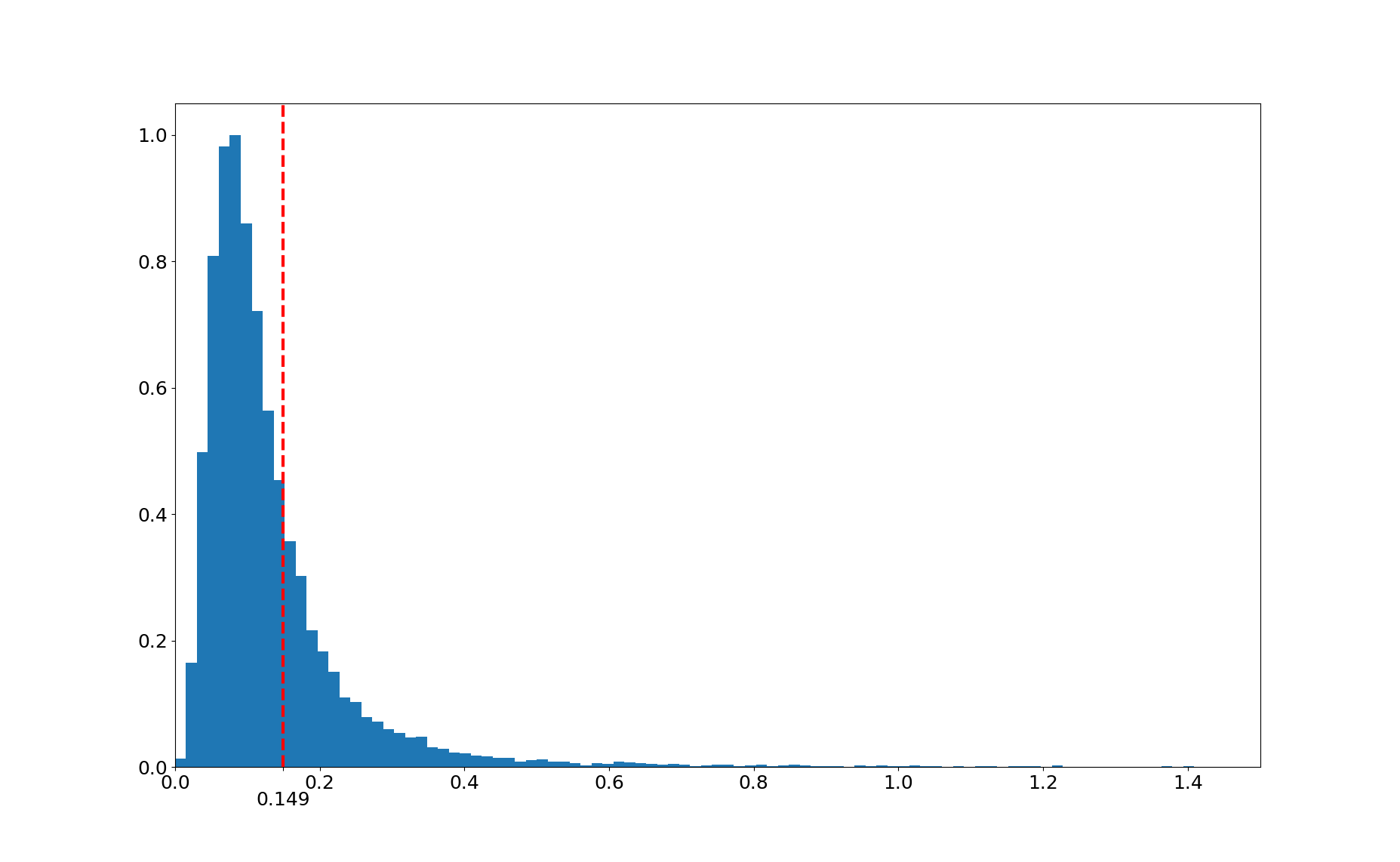}
        \label{fig:bunny_pdf_naive}
    \end{minipage}
    \begin{minipage}[c]{.22\linewidth}
        \centering
        \includegraphics[width=0.7\textwidth]{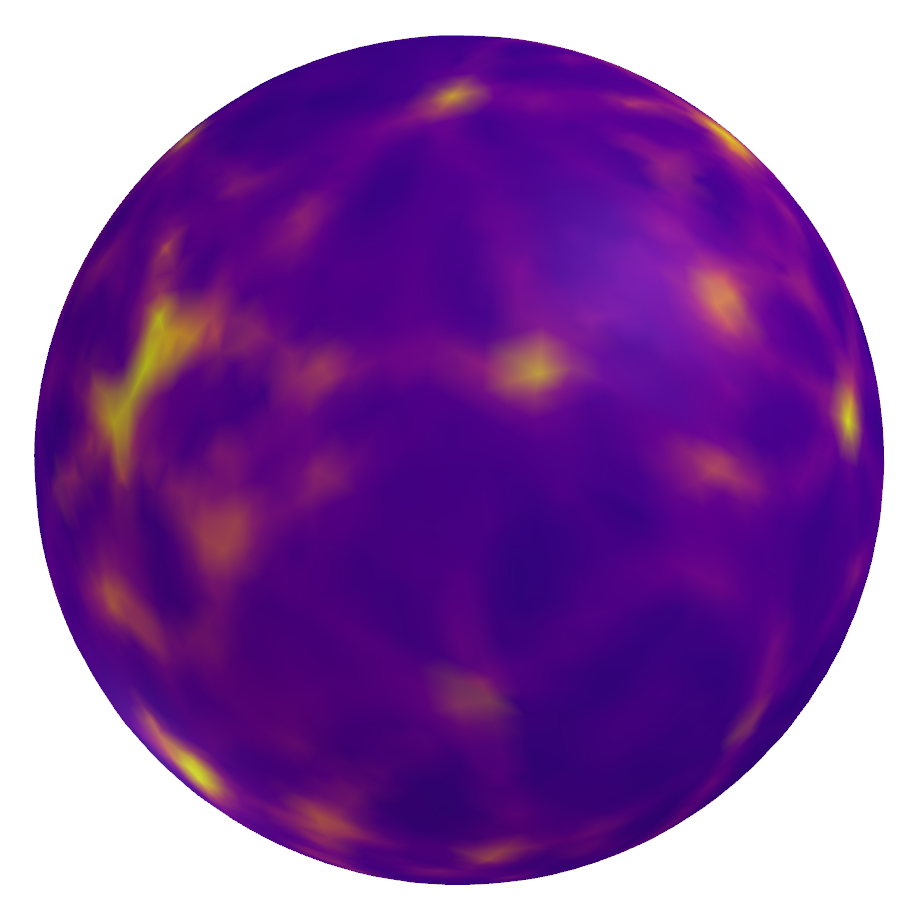}
        \par
        \includegraphics[width=\textwidth]{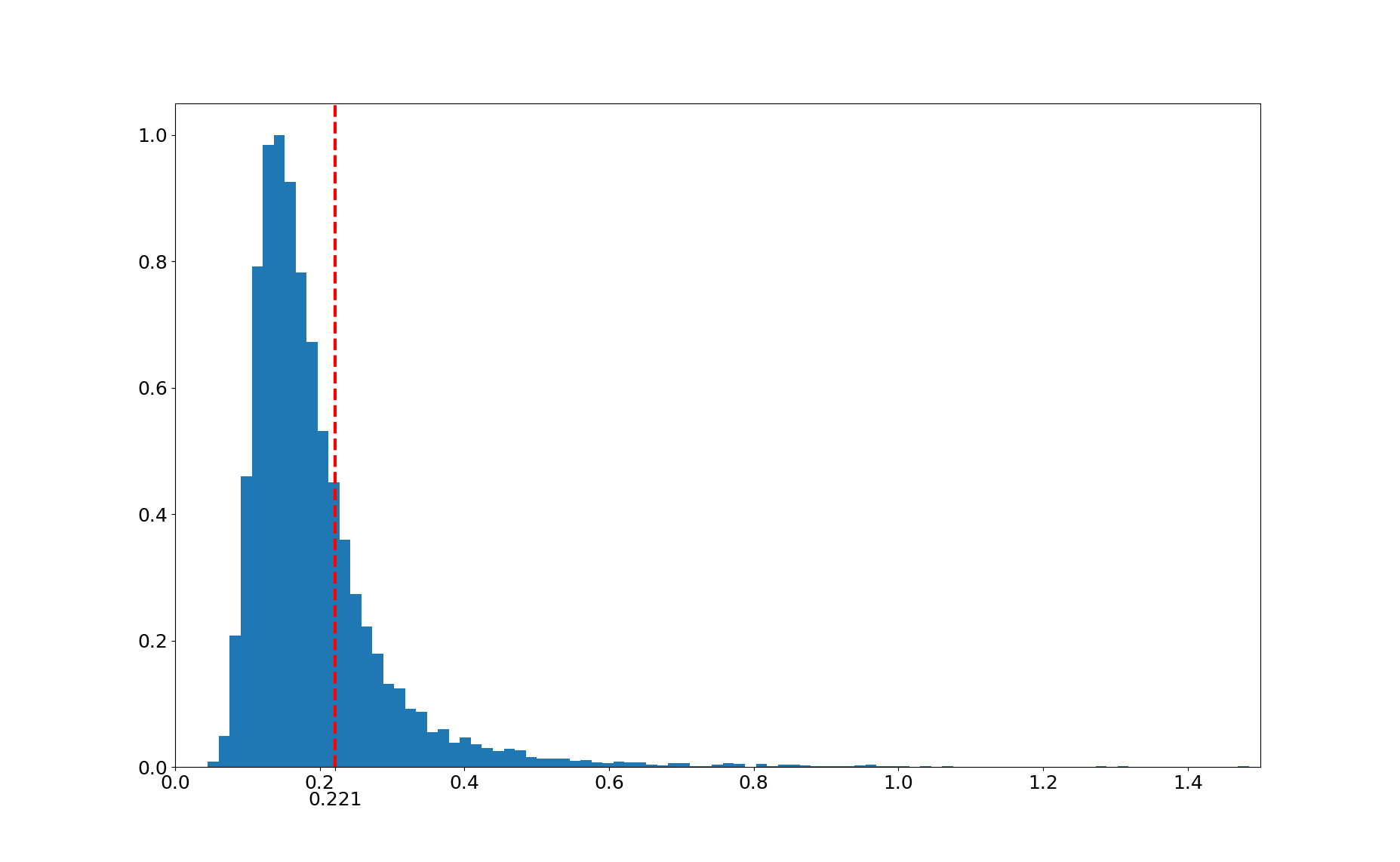}
        \label{fig:sphere_pdf_naive}
    \end{minipage}
    \begin{minipage}[t]{.10\linewidth}
        \centering
        \includegraphics[height=1.03in]{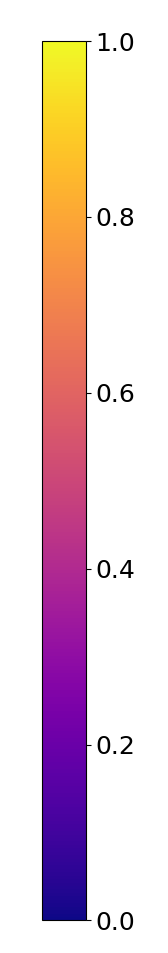}
        \label{fig:pdf_colorbar}
    \end{minipage}
    \caption{On the top row the estimated pdf is visualized with face colors. On the bottom row, the respective histograms of the estimated pdf values are shown. The dashed vertical red line indicates the value of the uniform pdf. For the two columns on the left, our proposed algorithm was used, while for the ones on the right, the naive approach.}
    \label{fig:pdf_estimation}
\end{figure*}

\subsection{Interaction Sampling} \label{sec:InteractionSampling}

We will now describe how we produce samples on the area of the surface that is affected by the brush. We begin by placing uniform samples on a disk tangent to the interaction point whose radius is the same as the brush's radius. We can, then, project these samples on the unaltered surface and then move them perpendicular to the tangent plane (at the interaction point) a distance equal to the brush function's value. Theoretically, for what we discussed in the previous section to be applicable this needs to be a parallel projection, however, we use the same procedure that we described in Section \ref{sec:SurfaceSampling} so that we don't affect the surface in a greater area than intended. The above process is shown in Figure \ref{fig:interaction_sampling} Next we compute the normals at the sampled position.

Suppose that $f$ is a 2D function defined over a 3D plane (its gradient $\nabla f$ then lies on the plane), and $\bm{n}$ is the normal vector to that plane (analogous to the z axis), then the following is an unnormalized vector, perpendicular to the graph of the function:
\begin{equation}
    - \nabla f + \bm{n}
    \label{eq:normal_from_gradient}
\end{equation}

\begin{table}[b]
    \centering
    \begin{tabular}{|l||c|c|}
        \hline
        \multirow{2}{*}{Shape} & \multicolumn{2}{c|}{Chamfer Distance $\times 10^3$ ($\downarrow$)} \\
        \cline{2-3}
        & Without WN & With WN \\
        \hline
        Bunny   & 9.021 & \textbf{8.995} \\
        Frog    & 8.095 & \textbf{7.921} \\
        Bust    & 7.167 & \textbf{7.139} \\
        Pumpkin & 8.646 & \textbf{8.506} \\
        Sphere  & 7.087 & \textbf{6.861} \\
        Torus   & 7.198 & \textbf{6.882} \\
        \hline \hline
        \textit{Average}    & 7.869  & \textbf{7.717} \\    
        \hline
    \end{tabular}
    \caption{Chamfer distances computed with 100000 points for weight normalization ablation.}
    \label{tab:wn_dist}
\end{table}

Accordingly, what we need in order to compute the normals of the samples is the gradient of the 2D function whose graph is the deformed surface. As we explained in the previous section, this function is the sum of the brush function and the function defined implicitly by the network. We can directly calculate the gradient of the brush function. By the implicit function theorem, the gradient of the implicit function is given by:
\begin{equation}
    - \frac{ \left( \nabla_{\bm{x}} f_\theta(\bm{x}) \right)_\parallel }{ \left( \nabla_{\bm{x}} f_\theta(\bm{x}) \right)_\perp }
    \label{eq:gradient_from_normal}
\end{equation}

\noindent where $\parallel$ denotes the component parallel to the tangent plane and $\perp$ the component perpendicular to it.

We, firstly, sample the model and discard the samples that lie inside a sphere centered at the interaction point with a radius equal to the brush's radius. Then, we sample the interaction. The number of samples we take is the number of discarded samples multiplied by an integer factor bigger than 1. We utilize this factor in order to balance the contributions of the model and interaction samples. We want the influence of the interaction samples to be larger since it is where the surface must change, while adapting to the size of the affected area. We use a factor of 10 for our experiments. Finally, the set of samples for the training is the union of the non-discarded model samples and the interaction samples.

\begin{table*}[t]
    \centering
    \begin{tabular}{|l||c|c|c|c|c|c|}
        \hline
        \multirow{3}{*}{Shape} & \multicolumn{6}{c|}{Mean Chamfer Distance $\times 10^3$ ($\downarrow$)} \\
        \cline{2-7}
        & \multicolumn{3}{c|}{Over whole surface} & \multicolumn{3}{c|}{Inside interaction area} \\
        \cline{2-7}
        & Ours & Naive & Simple Mesh & Ours & Naive & Simple Mesh \\
        \hline
        Bunny   & \textbf{9.407} & 14.106 & 11.127 & \textbf{5.527} & 12.919 & 17.707 \\
        Frog    & \textbf{8.172} & 12.865 & 8.756 & \textbf{4.750} & 9.805 & 17.051 \\
	    Bust    & \textbf{7.279} & 11.486 & 7.901 & \textbf{3.818} & 8.779 & 14.926 \\
	    Pumpkin & \textbf{8.774} & 13.558 & 11.489 & \textbf{4.315} & 5.910 & 20.693 \\
	    Sphere  & \textbf{7.209} & 12.550 & 7.399 & \textbf{3.555} & 6.117 & 12.982 \\
	    Torus   & \textbf{7.142} & 13.516 & 7.415 & \textbf{3.574} & 5.980 & 13.402 \\
        \hline \hline
        \textit{Average}    & \textbf{7.997} & 13.014 & 9.015 & \textbf{4.257} & 8.252 & 16.127 \\
        \hline
    \end{tabular}
    \caption{Comparison of our editing method with and without model samples (Ours and Naive, respectively) and direct mesh editing on a mesh with equivalent size (Simple Mesh). Chamfer distances are computed with 100000 points. The mean for each shape is taken over 10 independent edits.}
    \label{tab:mesh_comp}
\end{table*}

\section{Experiments} \label{sec:Experiments}

In this section, we qualitatively and quantitatively evaluate our proposed 3DNS in various 3D objects and under several different surface edits. For additional results and visualizations, please refer to the Supplementary Material section below. Also, the source code and a demo video can be found at the paper's webpage (\url{https://pettza.github.io/3DNS/}).

\subsection{Dataset} 

For our experiments we have created a small dataset of six 3D shapes, see Fig.~\ref{fig:meshes}. For four of them, we start from a  mesh representation: The Stanford Bunny, which comes from \cite{turk_zippered} and the frog, pumpkin, and bust, which come from TurboSquid \footnote{\url{https://www.turbosquid.com}}. The other two shapes are a sphere and a torus which are usually provided by 3D modeling software as starting shapes. We start from an analytical representation of these shapes. Specifically, we use a sphere with a radius of 0.6 and a torus with a major radius of 0.45 and a minor radius of 0.25. For our implementation, we use PyTorch \cite{pytorch}. 
As a pre-processing step, we normalize the coordinates of the four meshes of our dataset by translating and scaling them uniformly so that they lie inside $[-(1 - b), 1 - b]^3$, where $b$ is a positive number. The latter parameter is used so that there is space around the models where we can edit them. We set $b$ to 0.15.
We train networks to represent the shapes of the dataset by sampling them uniformly. The architecture of the networks is SIREN with 2 hidden layers and 128 neurons each, and weight normalization (except in Section \ref{sec:WeightNormAbl}). We use 120000 samples for the loss of equation \ref{eq:sdf_loss} and another 120000 for the losses of equations \ref{eq:eikonal_loss} and \ref{eq:empty_space_loss} and train for $10^6$ iterations.

\subsection{Performance Metric}

For quantitative comparisons, we use the Chamfer distance, which is a common metric used for geometric tasks. If $A$, $B$ are two point clouds, their Chamfer distance is defined as follows:
\begin{equation}
    \mathit{CD}(A, B) = \frac{1}{|A|} \sum_{a \in A} \min_{b \in B} d(a, b) + \frac{1}{|B|} \sum_{b \in B} \min_{a \in A} d(a, b) 
\end{equation}

\subsection{Brush Parameters} \label{sec:BrushParams}

We demonstrate how the brush parameters allow the user to control the edit in Figure \ref{fig:brush_params}. Specifically, we use different radii and intensities on the same interaction point on the sphere. For each row, the intensity is the same and is shown on the left. Similarly, for each column, the radius is the same and is shown on the top.

\subsection{Weight Normalization Ablation Study} \label{sec:WeightNormAbl}

\newcommand{\meshcompexfig}[1]{%
   \includegraphics[trim=550 450 960 220, clip, width=\linewidth]{Figures/mesh_comparison_example/#1.png}%
}

\begin{figure}[b]
    \centering
    \begin{subfigure}{0.22\linewidth}
        \meshcompexfig{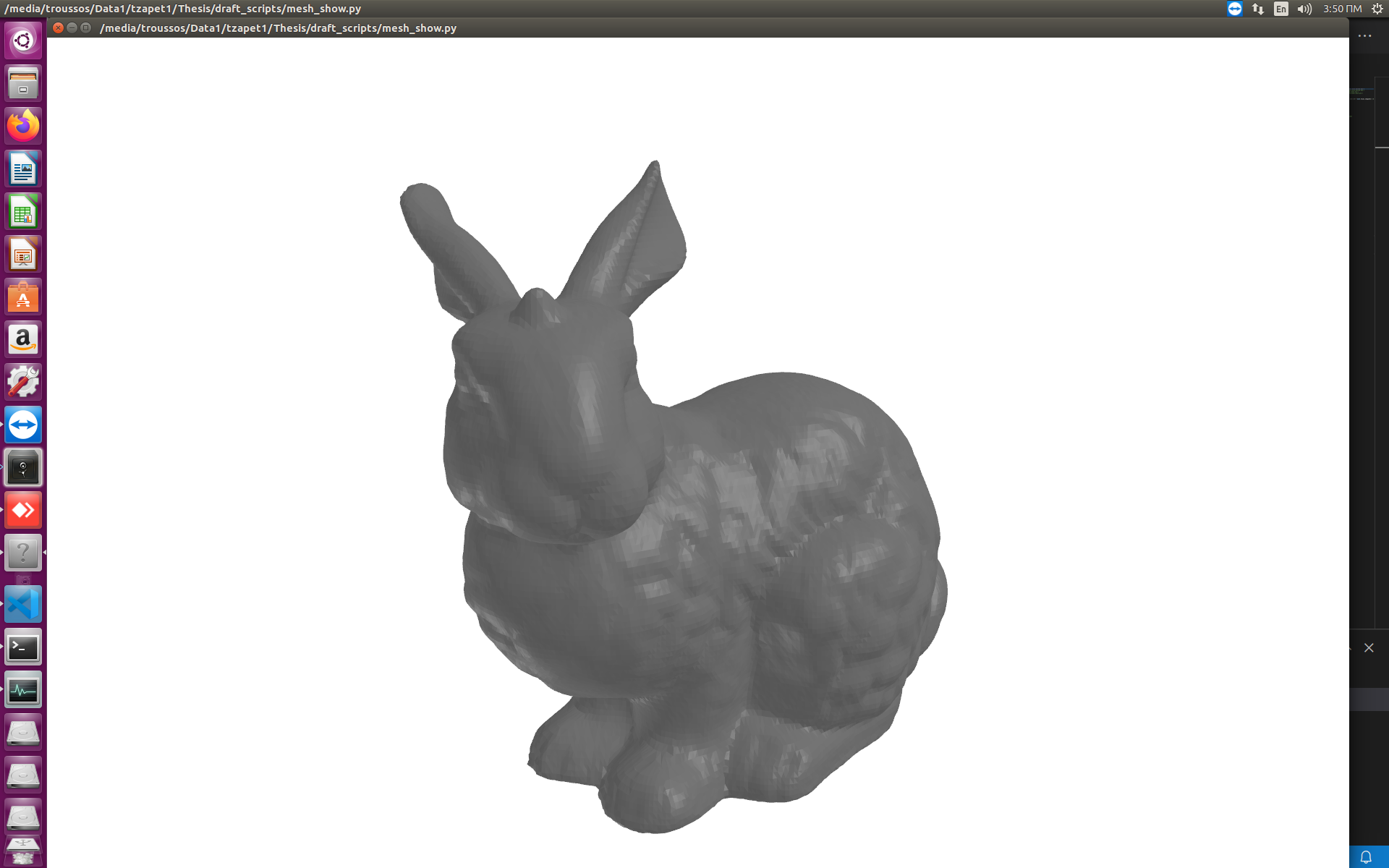}
    \end{subfigure}
    \hfill
    \vrule
    \hfill
    \begin{subfigure}{0.22\linewidth}
        \meshcompexfig{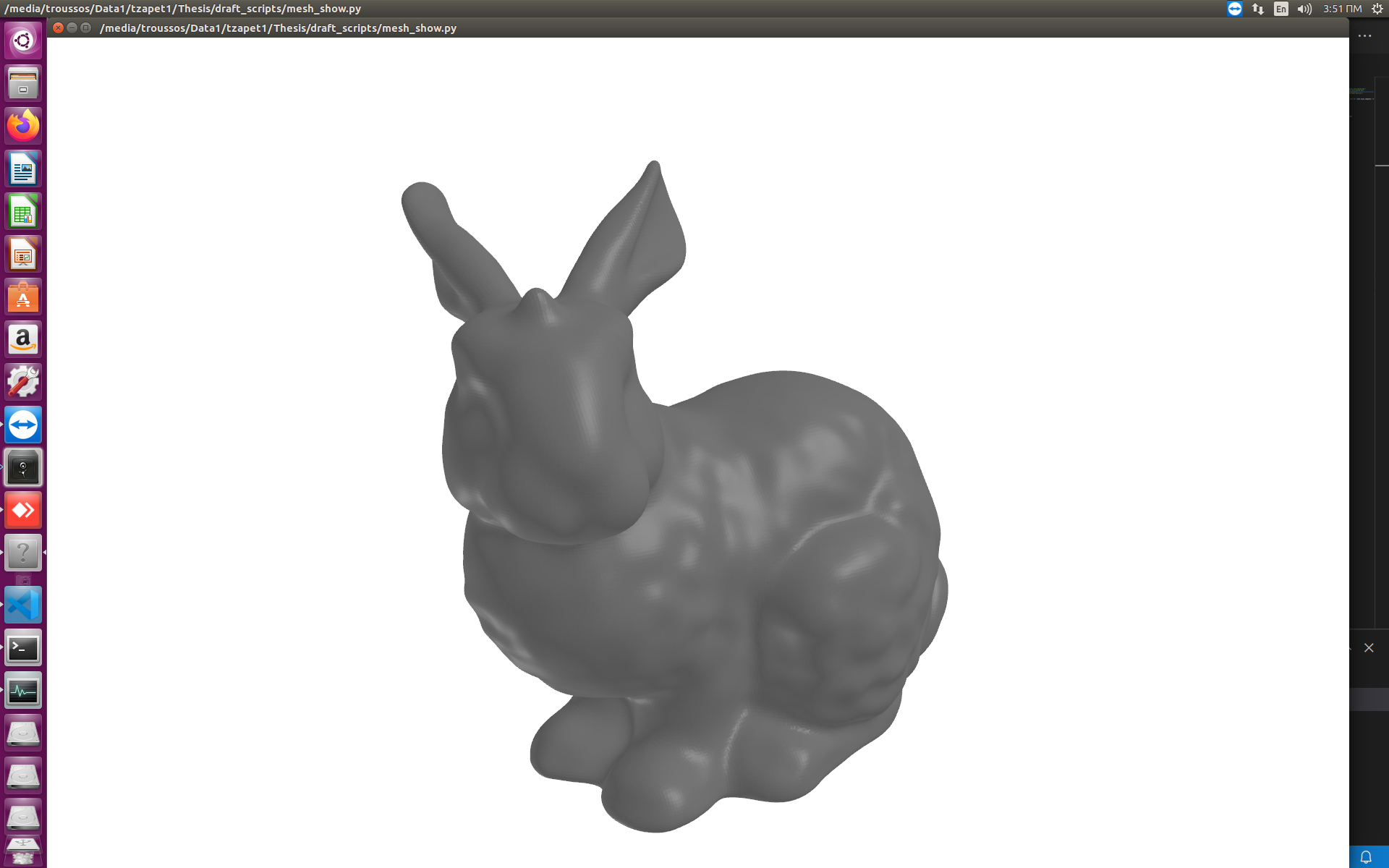}
    \end{subfigure}
    \begin{subfigure}{0.22\linewidth}
        \meshcompexfig{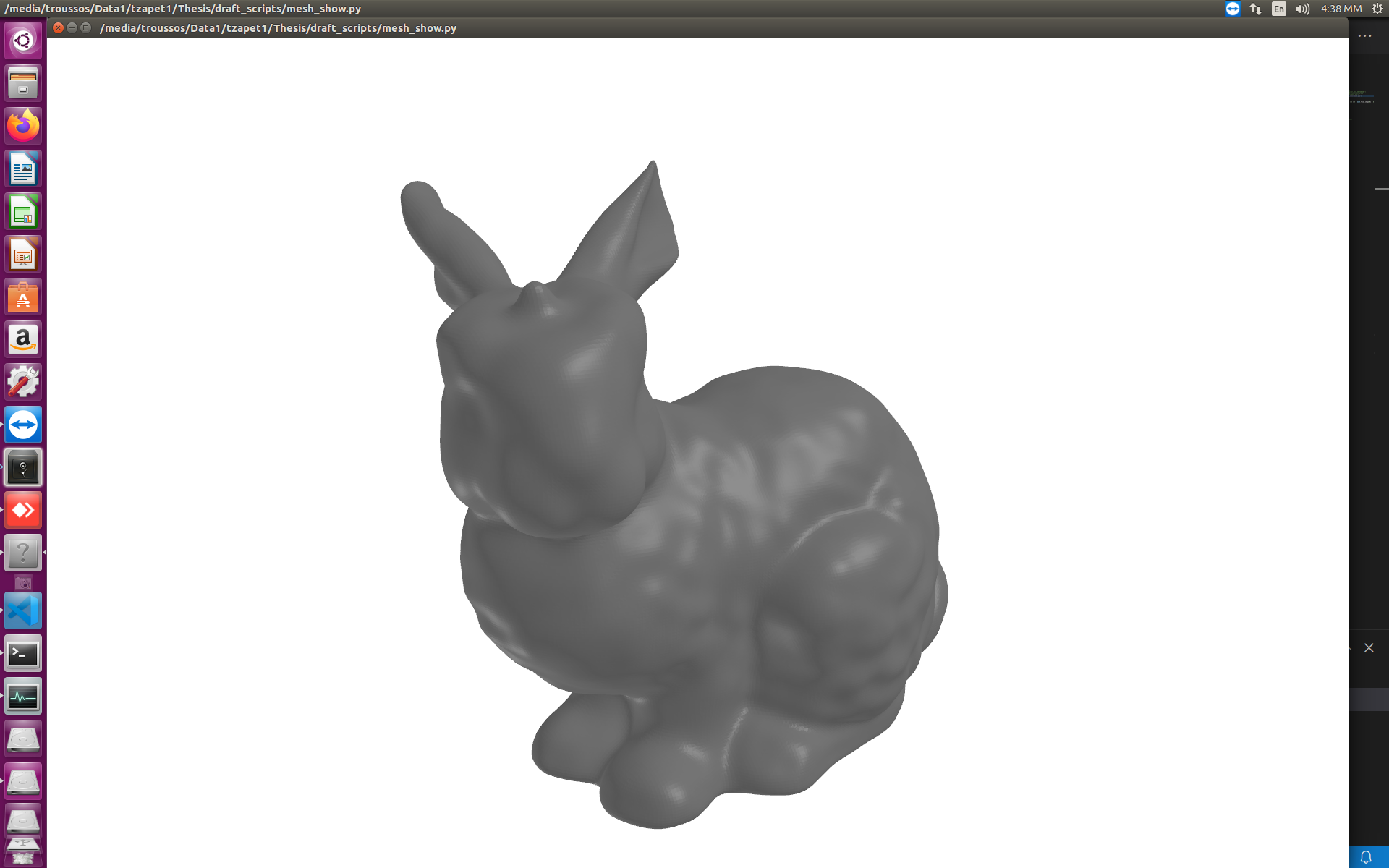}
    \end{subfigure}
    \begin{subfigure}{0.22\linewidth}
        \meshcompexfig{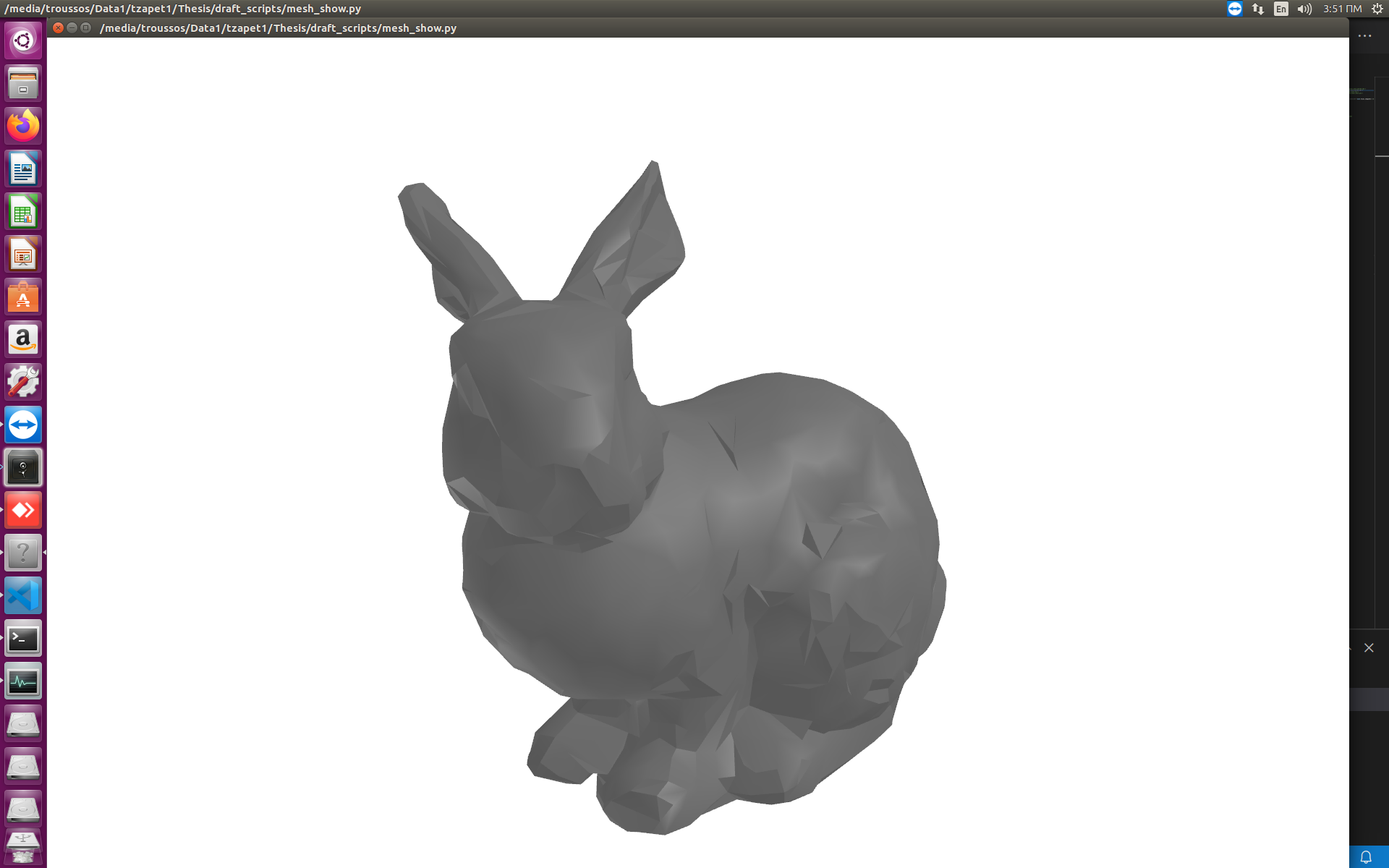}
    \end{subfigure}
    \caption{Example of an edit using different methods. From left to right, ground truth (ideal edit result), ours, naive and simplified mesh.}
    \label{fig:example_edit}
\end{figure}

In order to prove the improvements provided by weight normalization, we train six models for the shapes we presented above in the first case with weight normalization and in the second case without. The Chamfer distance between a point cloud sampled from the ground truth models and one sampled from the trained network by our sampling algorithm is given in Table \ref{tab:wn_dist}. It can be seen that, even though not by a large margin, the models trained with weight normalization perform better in every case.

\subsection{PDF Estimation} \label{sec:PDFEstimation}

We want to study the uniformness of the stationary distribution of our sampling algorithm. We do this by estimating the pdf over the surface. Firstly, we create a triangle mesh of the surface using the Marching Cubes algorithm \cite{marching_cubes}. We, then, generate samples with our algorithm for $N$ iterations, with $M$ samples per iteration. In effect, we are simulating $M$ Markov chains. For each triangle of the mesh, we count the number of samples $c$ that are closest to it. The estimated mean value of the pdf over the triangle is then:
\begin{equation}
    pdf = \frac{c}{N \cdot M \cdot A}
    \label{eq:pdf_estimation}
\end{equation}

\noindent where $A$ is the area of the triangle.

We visualize the estimated pdf for the bunny and the sphere with the face colors in Figure \ref{fig:pdf_estimation}. In the same figure, we show histograms of the pdf estimations, as well. The value of the uniform pdf, which is equal to the inverse of the surface area, is shown in the histograms with a dashed vertical red line. We can see that the histograms are centered tightly around this value, indicating that the stationary distributions of our sampling process are quite uniform. For comparison, we give the corresponding results for the naive sampling outlined in Section \ref{sec:SurfaceSampling}, in the same Figure. Here, we notice the bright areas which are sampled more frequently than the rest of the surface, as well as the longer tails of the histograms and the fact that they are off-centered.

\subsection{Mesh Editing Comparison} \label{sec:MeshComp}

We compare our proposed method with the editing of meshes which is by far the most popular representation for 3D modeling and sculpting and, also, the naive approach of using only interaction samples. We edit a mesh by changing the positions of the vertices that lie inside the interaction area (the sphere that was used to discard samples). We follow the process described in Section \ref{sec:InteractionSampling}, the only difference being that, instead of projecting samples from the tangent plane onto the surface, we compute a vertex's corresponding position on that plane as the intersection of a ray starting from the vertex with direction along its normal. In order to have a fair comparison, we use a mesh with approximately the same size as the network. We begin with a high-resolution mesh as the ground truth (this is the mesh the network was trained on for the four meshes and one constructed via Marching Cubes \cite{marching_cubes} for the sphere and torus) and, following \cite{nglod}, use quadratic decimation \cite{mesh_decimation} to get the smallest mesh with size larger or equal to the network's size. Afterward, we perform the same ten edits on these three representations. Each edit is performed on the unedited models. We compute the mean Chamfer distance of the network and the simplified mesh to the ground truth mesh, over the whole surface, as well as only inside the interaction areas. We set the brush's radius to 0.08 and its intensity to 0.06 for all the edits. The results are summarized in Table \ref{tab:mesh_comp}, where it is shown that our method outperforms the other approaches. We, also, provide an example in Figure \ref{fig:example_edit}.

\section{Limitations and Future Work} \label{sec:Limitations}

Despite the successes of neural SDFs and the results we present above, there are also drawbacks. One problem with this way of editing is that the shape cannot easily be edited outside the bounding box where the eikonal loss has been applied because there the network does not approximate an SDF. Also, since the neural network function is smooth it is difficult to model hard edges and corners using a neural SDF.
However, there is research on neural SDFs that use auxiliary data structures to represent the surface locally \cite{deepls, instant-ngp, nglod} that can be potentially utilized to address these shortcomings, which is a direction we aim to pursue. Furthermore, our framework could in the future be extended in editing neural scene representations such as NeRFs \cite{mildenhall2020nerf}. 

\section{Conclusion}

We have presented 3DNS, a method for editing neural SDF representations in an interactive fashion inspired by existing 3D software. We believe that research towards the editability of these representations can help render them viable for more applications, either scientific or artistic in nature. Nevertheless, in order to compete with the existing tools and capabilities of meshes, more research is required.

{\small
\printbibliography
}

\clearpage

\renewcommand\appendixpagename{Supplamentary Material}

\appendixpage

\begin{appendices}

\begin{refsection}

\section{Other Brush Types}

In the main paper (Section \ref{sec:Brushes}) we used a single brush template, which was based on a quintic polynomial. Here, we describe a more general approach.

\subsection{Smoothstep Functions}

In computer graphics, the need to transition smoothly from one real number to another arises very frequently. For this purpose, various functions are used, which take the value $0$ for $x < 0$, the value $1$ for $x > 1$, and go from $0$ to $1$ in the interval $\left[0, 1\right]$ in an continuously differentiable increasing manner, with vanishing derivatives at $0$ and $1$. In graphics, such a function is usually called \emph{smoothstep}. Refer to Inigo Quilez's site \cite{quilez_smoothstep} for a presentation of some smoothstep functions.

\subsection{Radially Symmetric Brushes}

If $f$ is a smoothstep function, then we can define a brush template that is radially symmetric, as follows:

\begin{equation}
    b_T(\bm{x}) = f\left( 1 - \| \bm{x} \| \right)
\end{equation}

Any function from \cite{quilez_smoothstep} can be used. The brush we use for our experiments is of this type. Besides smoothsteps, any function $f$ defined over $\left[0, 1\right]$ that has value $0$ at $x = 0$ and a maximum value of $1$ can be used in the equation above.

\subsection{Vector Brushes}

A possible extension we can make to our brush formulation is to allow the brush template function to take vector values instead of just scalars. Such an extension would allow to create brushes that twists the surface locally around the interaction point.

\section{Pre-training}

\begin{figure}[t]
    \centering
    \includegraphics[width=\linewidth]{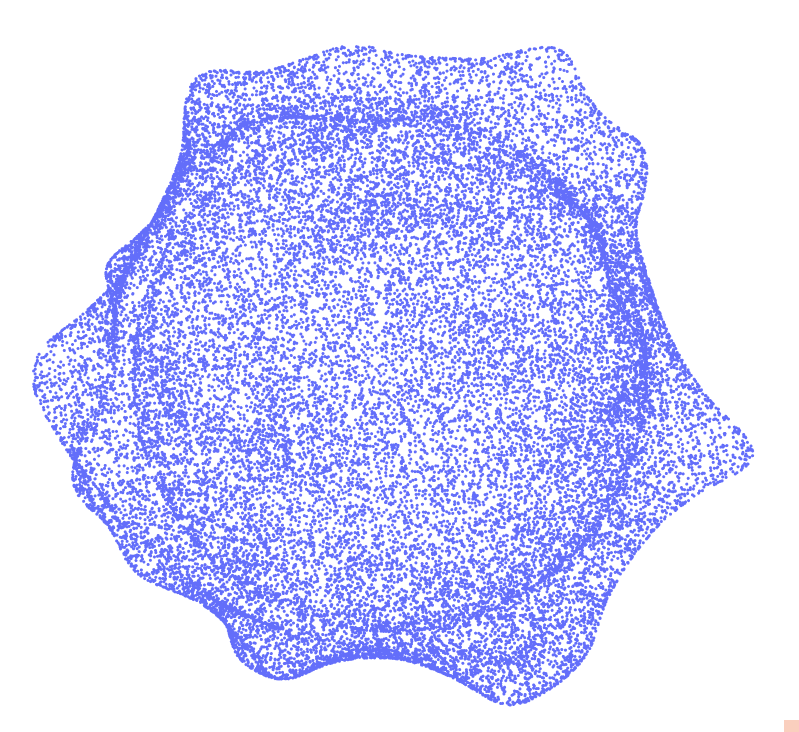}
    \caption{Sampled zero-level set of a neural network that was unsuccessfully trained to represent a sphere.}
    \label{fig:failcase}
\end{figure}

During the initial training of the networks so that they represent the unedited shapes we encountered a problem. Sometimes the training fails. Figure \ref{fig:failcase} shows the sampled zero-level set of a network that was (unsuccessfully) trained to represent a sphere. A sphere can actually be discerned. Nevertheless, there also exists an outer shell. This situation manifested for different shapes as well. We hypothesize that this is due to the nature of the loss function, the network's size, and poor initialization of the network weights.

\newcommand{\pdffig}[1]{
    \begin{minipage}[c]{.14\linewidth}
        \centering
        \includegraphics[width=0.7\textwidth]{Figures/supplamentary/pdf_estimation/#1_pdf_ours_mesh.png}
        \par
        \includegraphics[width=\textwidth]{Figures/supplamentary/pdf_estimation/#1_pdf_ours_hist.png}
        \par
        \includegraphics[width=0.7\textwidth]{Figures/supplamentary/pdf_estimation/#1_pdf_naive_mesh.png}
        \par
        \includegraphics[width=\textwidth]{Figures/supplamentary/pdf_estimation/#1_pdf_naive_hist.png}
        \label{fig:supp_#1_pdf}
    \end{minipage}
}

\begin{figure*}[t]
    \centering
    \pdffig{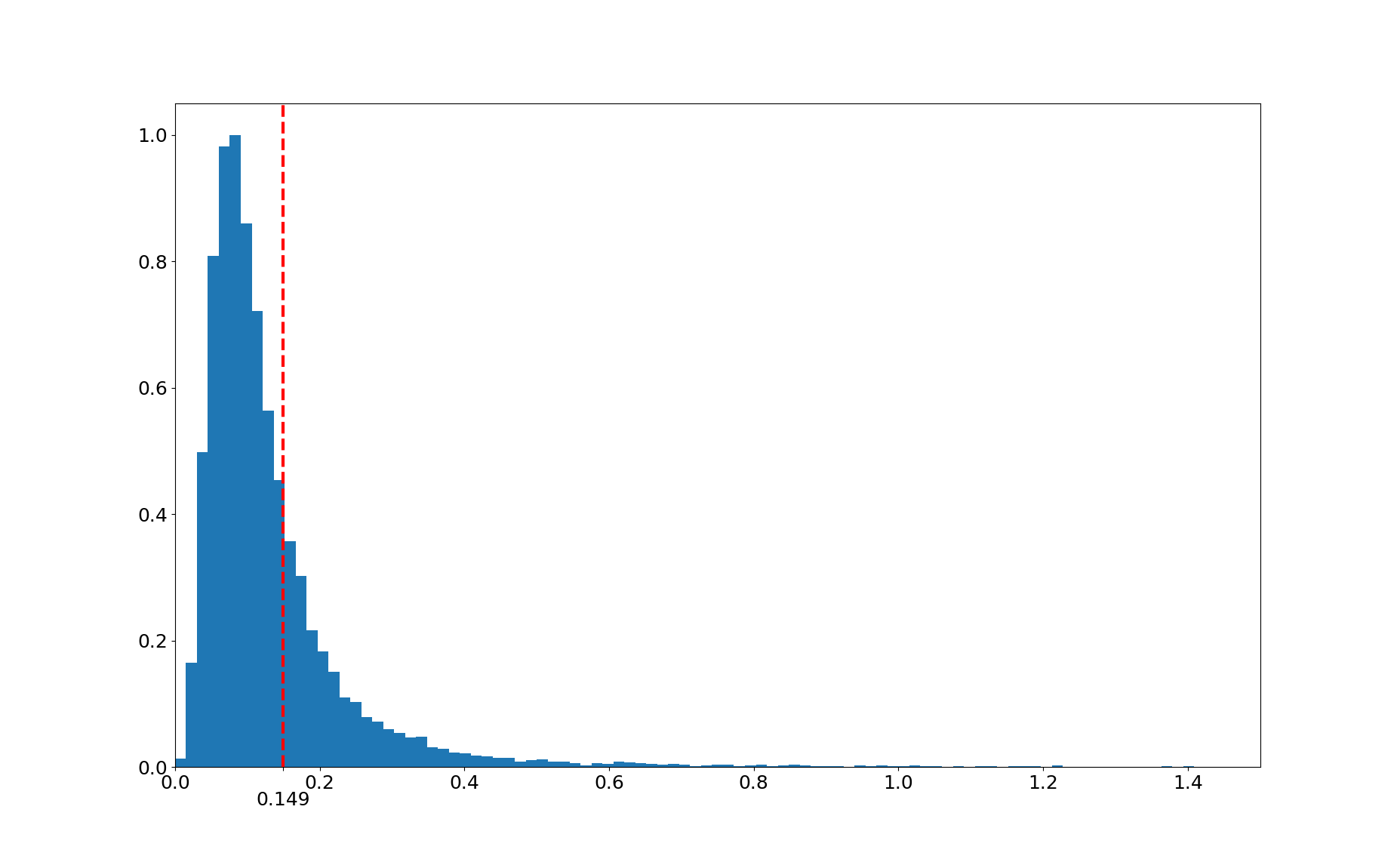}
    \pdffig{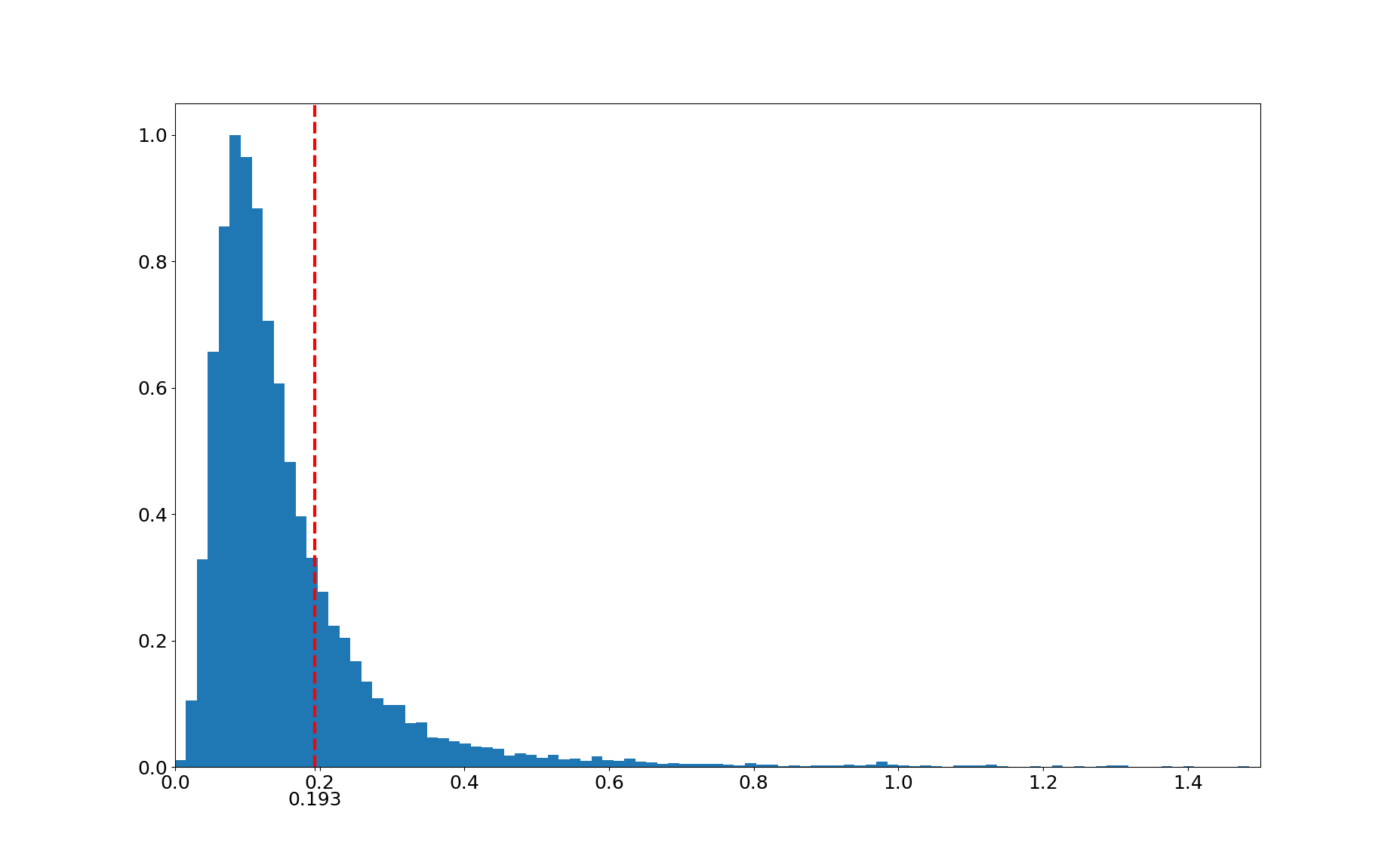}
    \pdffig{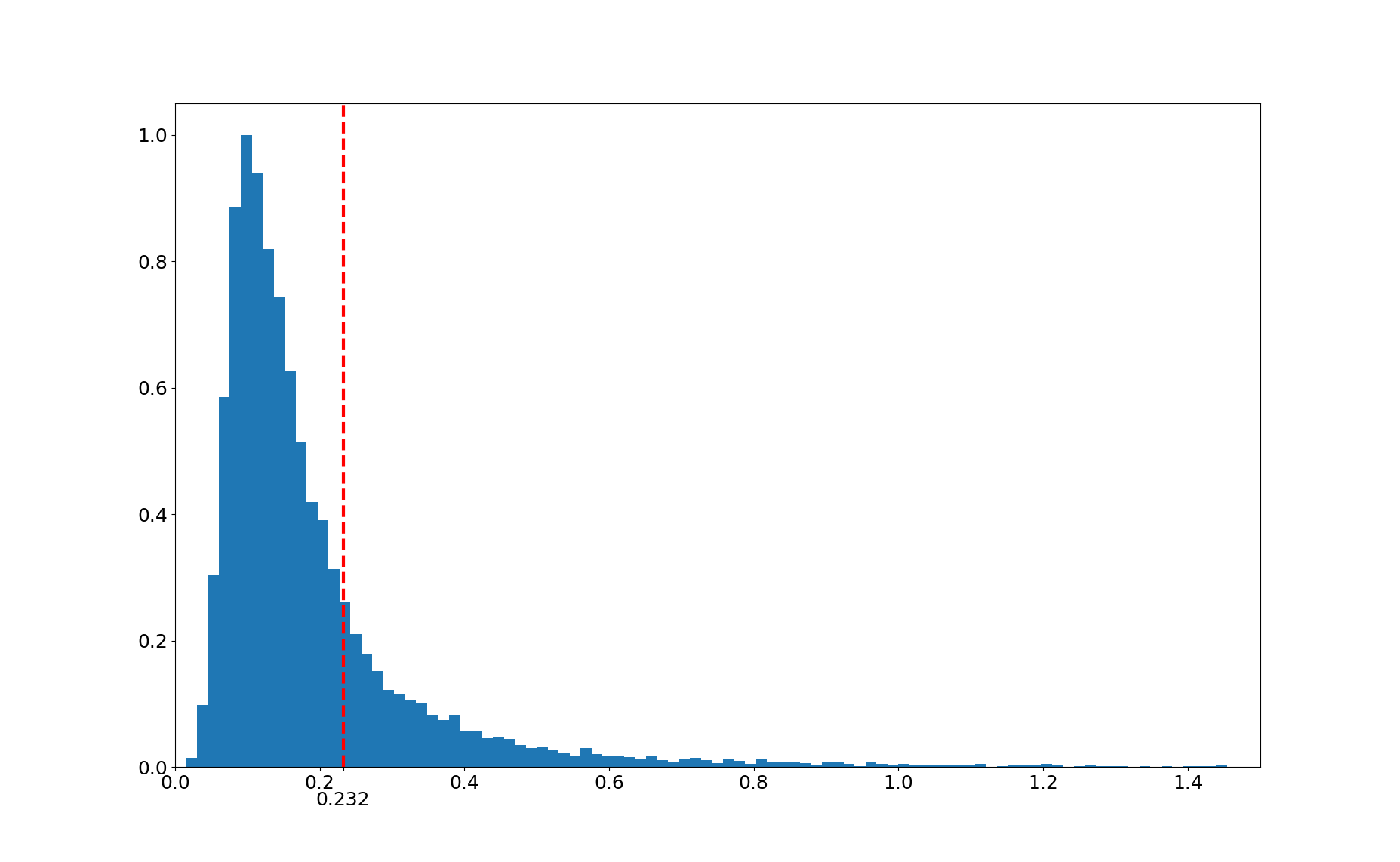}
    \pdffig{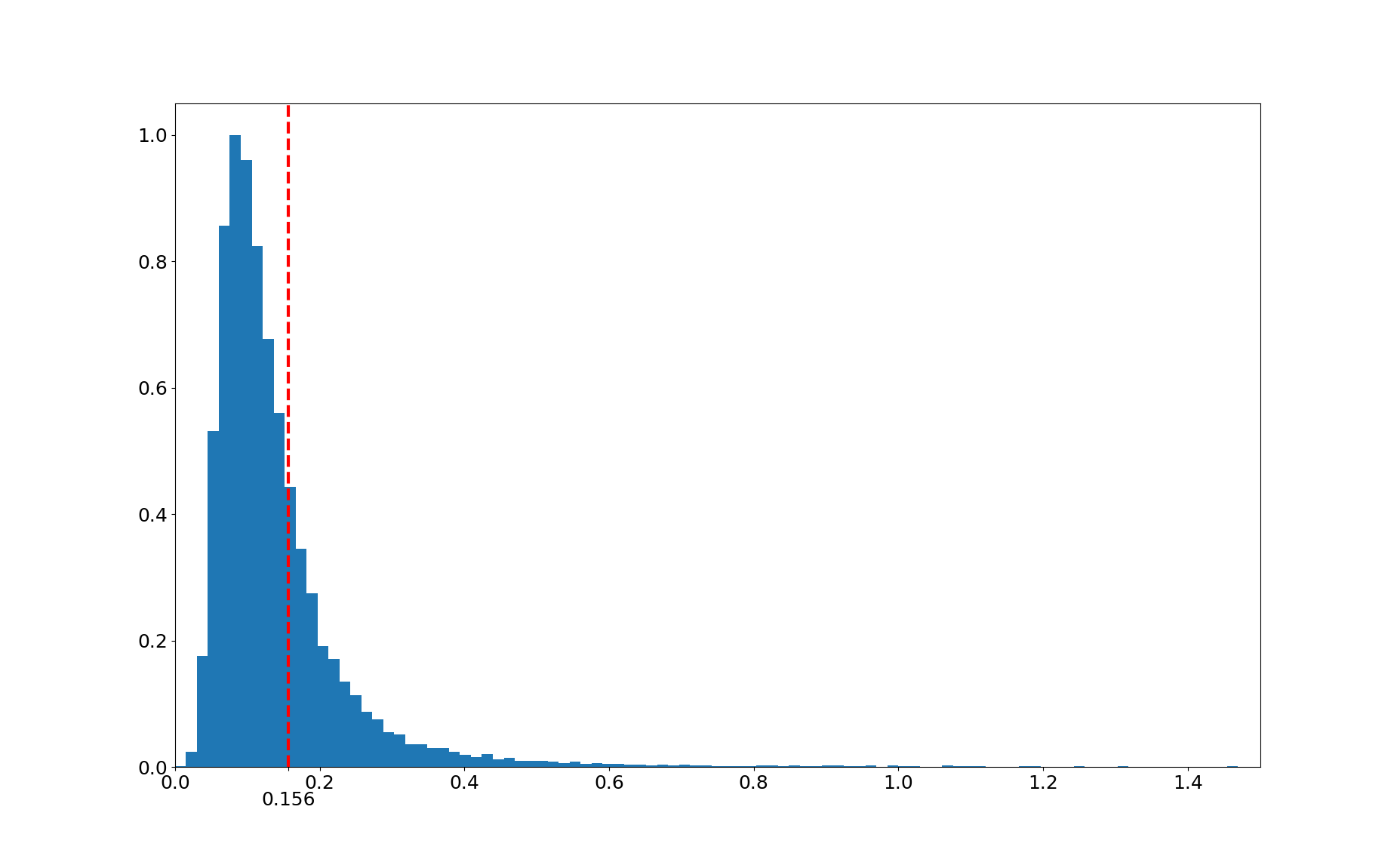}
    \pdffig{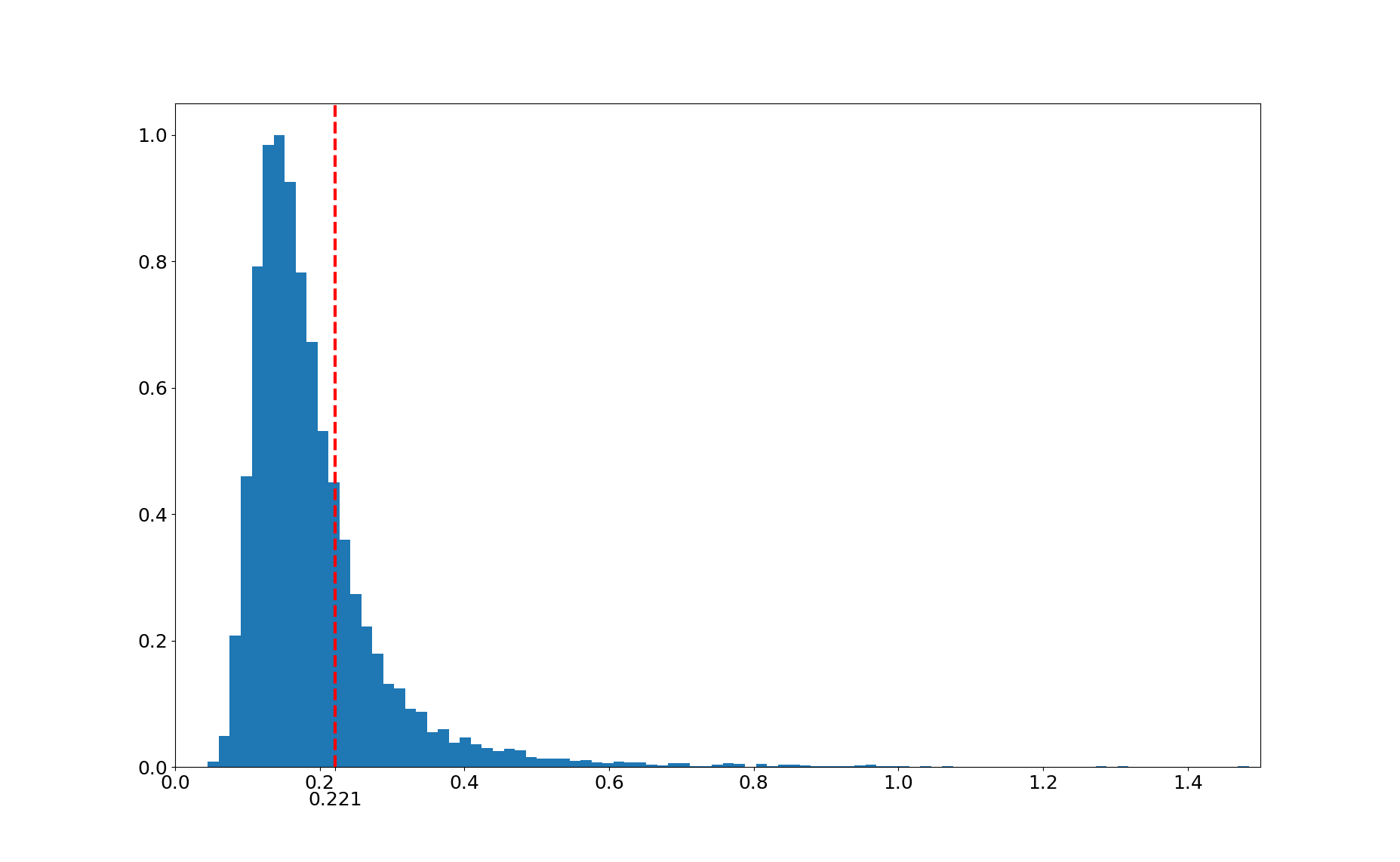}
    \pdffig{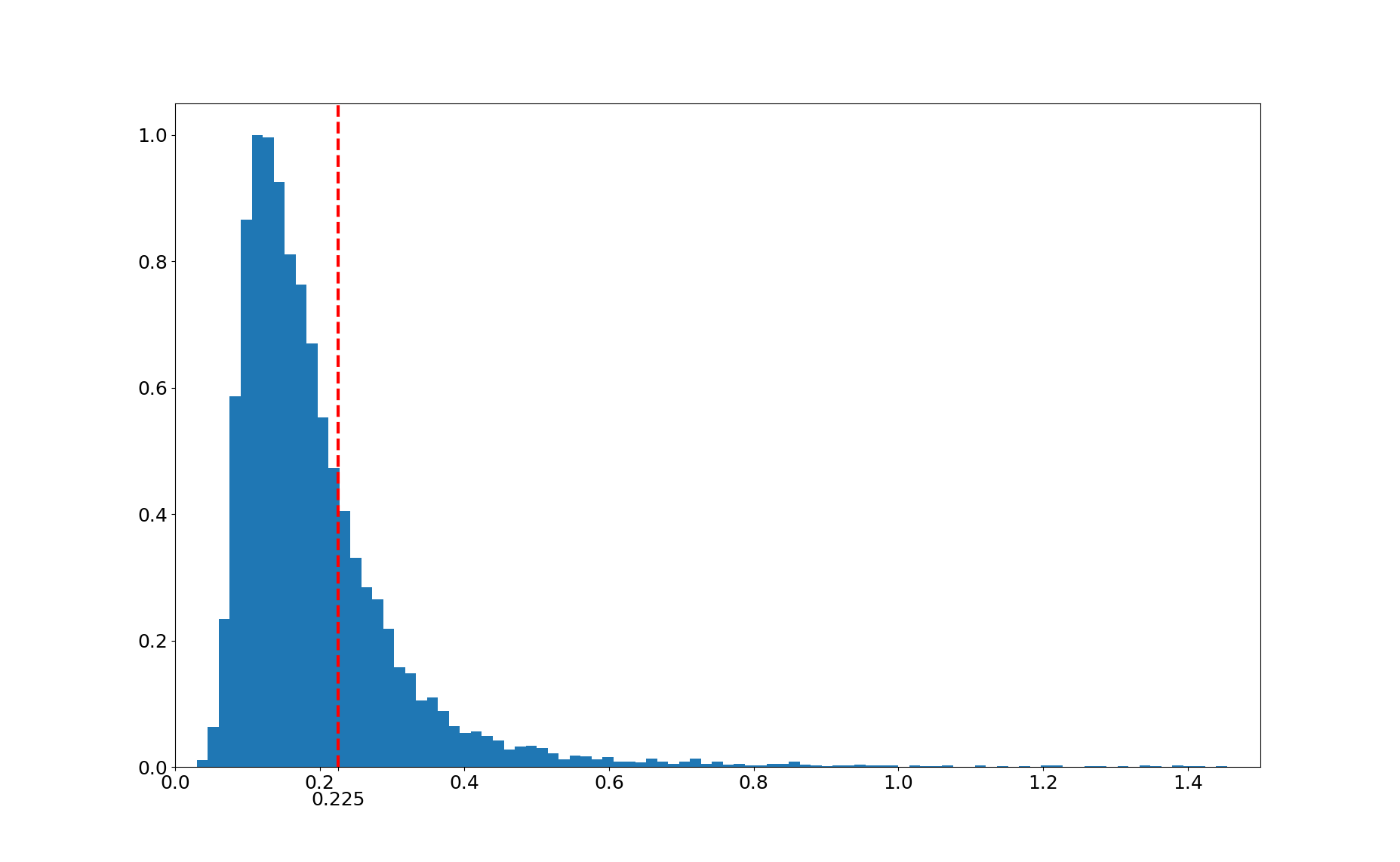}
    \begin{minipage}[c]{.04\linewidth}
        \centering
        \includegraphics[trim=50 0 10 0, clip, height=1.7in]{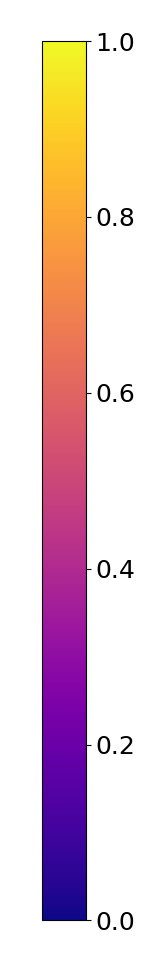}
        \label{fig:supp_pdf_colorbar}
    \end{minipage}
    \caption{The results of PDF estimation for all the shapes in the dataset. For the top two rows our proposed algorithm was used, while for the bottom two the naive approach. For an explanation of the figures refer to the main paper (Section \ref{sec:PDFEstimation}). Please zoom in for details.}
    \label{fig:supp_pdf_results}
\end{figure*}

\begin{figure*}[t]
    \centering
    \begin{subfigure}{0.19\linewidth}
        \centering
        \includegraphics[trim=0 10 0 20, clip, height=0.9in]{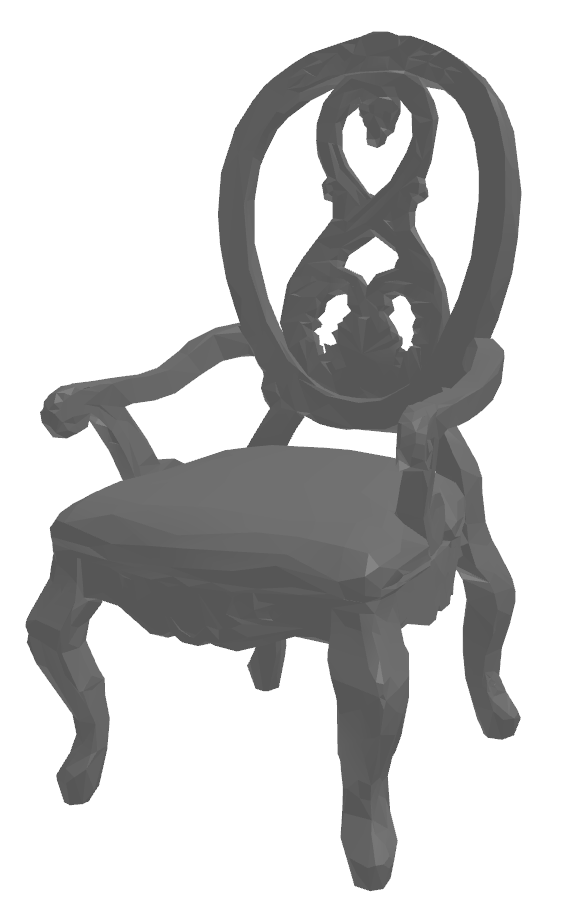}
        \caption{Dining chair}
        \captionsetup{aboveskip=0pt,font=it}
        \caption*{\tiny{4dd46b9657c0e998b4d5420f7c27d2df}}
        \label{fig:fancychair_shape}
    \end{subfigure}
    \begin{subfigure}{0.19\linewidth}
        \centering
        \includegraphics[height=0.9in]{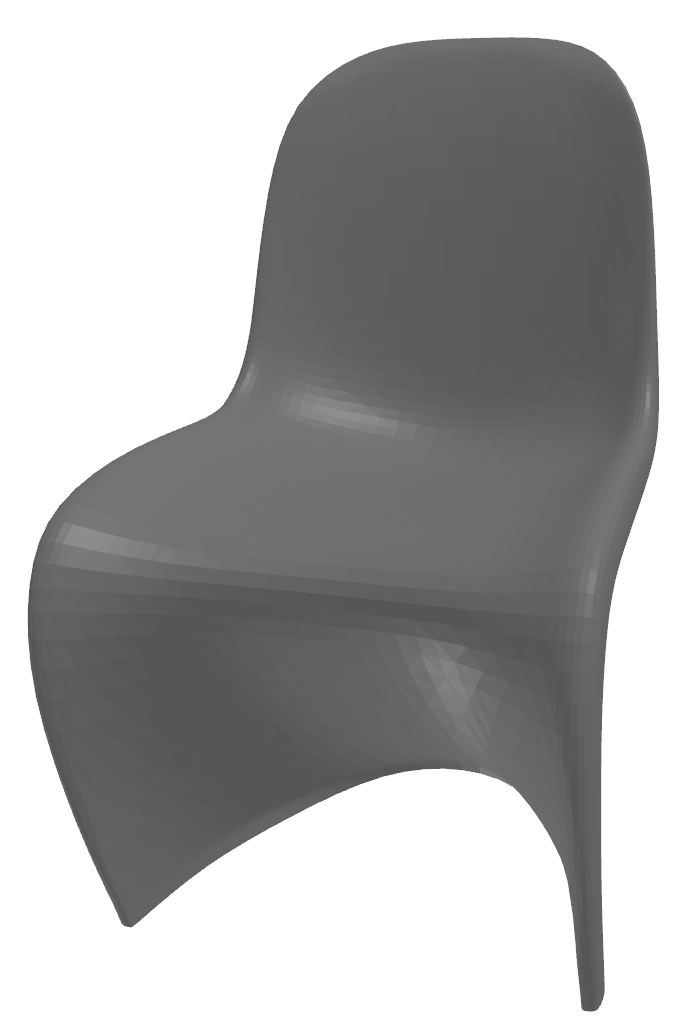}
        \caption{Chair}
        \captionsetup{aboveskip=0pt,font=it}
        \caption*{\tiny{02e76cb4f1039c482eb499cc8fbcd}}
        \label{fig:chair_shape}
    \end{subfigure}
    \begin{subfigure}{0.19\linewidth}
        \centering
        \includegraphics[height=0.9in]{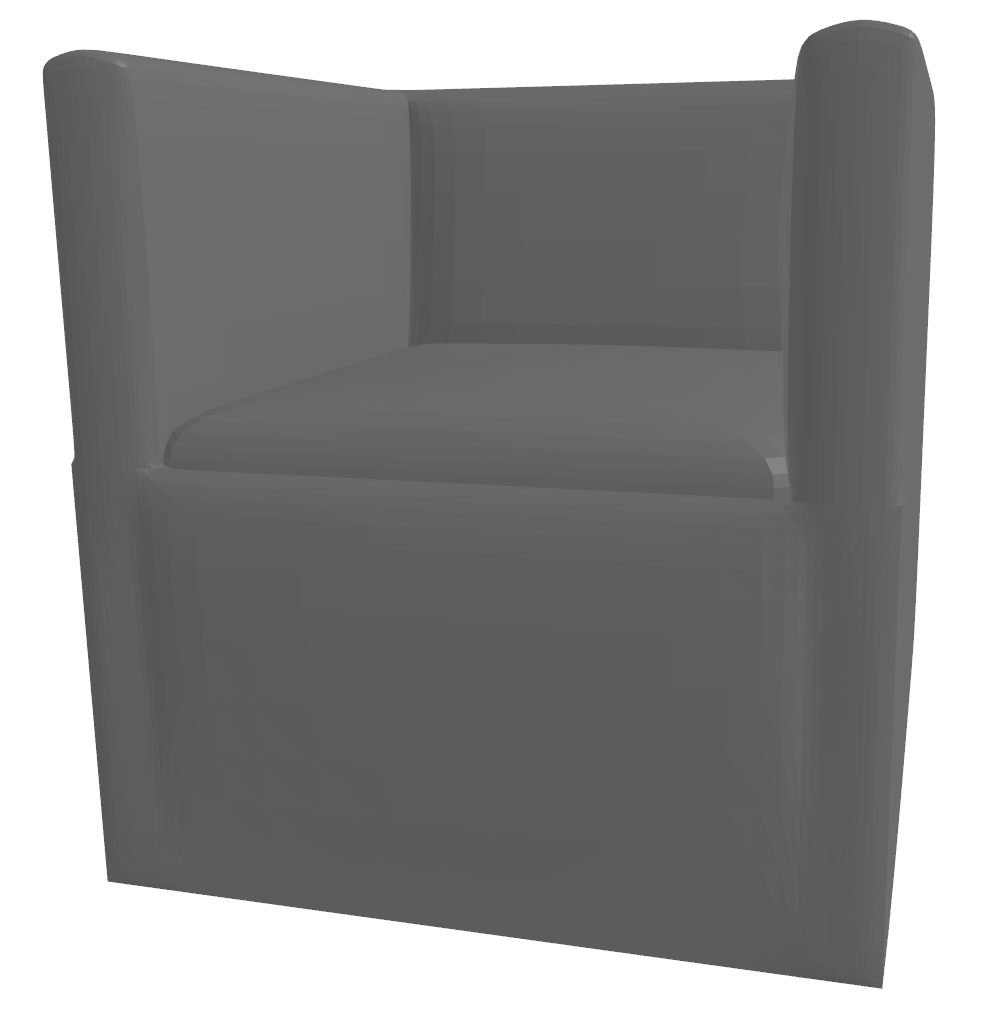}
        \caption{Armchair}
        \captionsetup{aboveskip=0pt,font=it}
        \caption*{\tiny{c5d880efc887f6f4f9111ef49c078dbe}}
        \label{fig:armchair_shape}
    \end{subfigure}
    \begin{subfigure}{0.19\linewidth}
        \centering
        \includegraphics[height=0.9in]{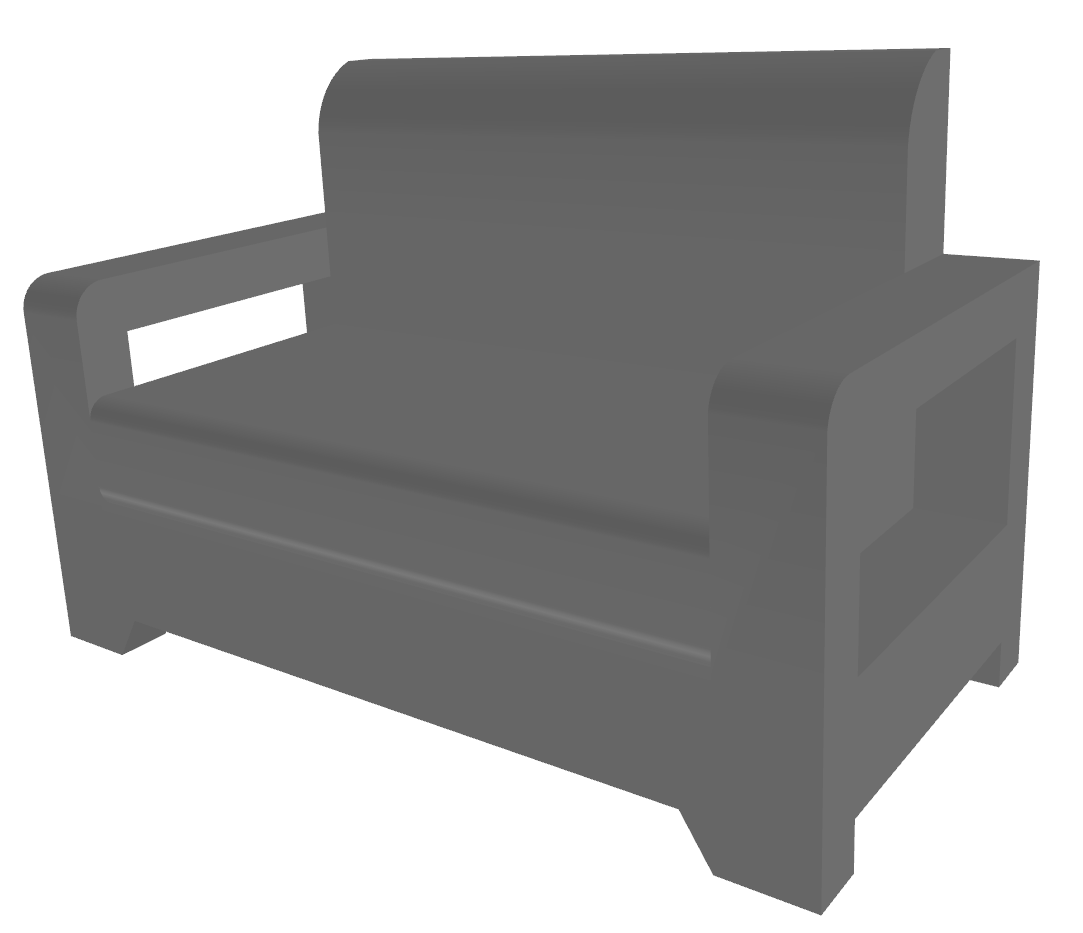}
        \caption{Sofa}
        \captionsetup{aboveskip=0pt,font=it}
        \caption*{\tiny{bcff6c5cb4127aa15e0ae65e074d3ee1}}
        \label{fig:sofa_shape}
    \end{subfigure}
    \begin{subfigure}{0.19\linewidth}
        \centering
        \includegraphics[height=0.9in]{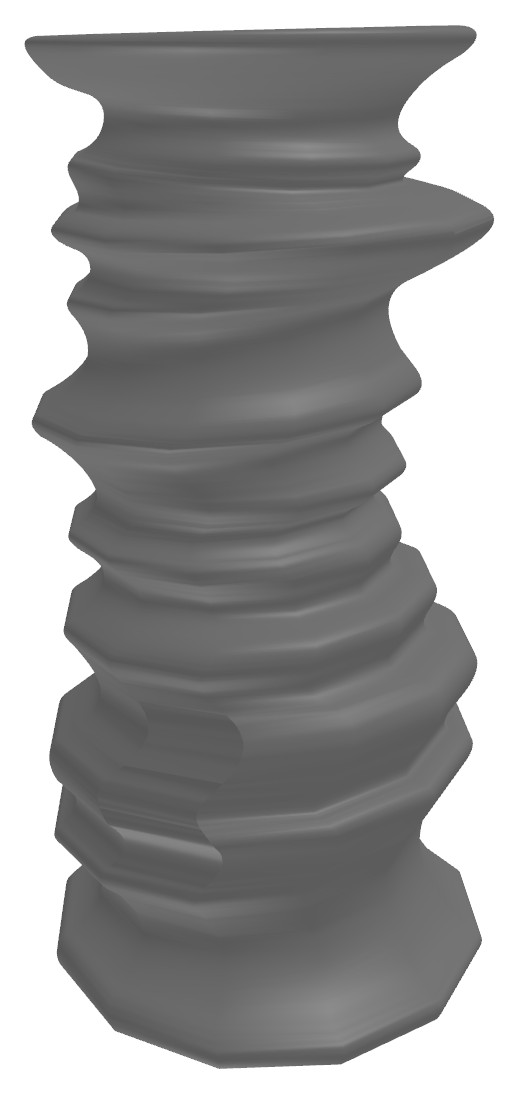}
        \caption{Vase}
        \captionsetup{aboveskip=0pt,font=it}
        \caption*{\tiny{13375f8fce3142e6597d391ab6fcc1}}
        \label{fig:vase_shape}
    \end{subfigure}
    \caption{ShapeNet shapes used for surface editing comparisons. Beneath each caption the ShapeNet model ID of the corresponding shape is written in italics.}
    \label{fig:shapenet_shapes}
\end{figure*}

The loss function (equations \ref{eq:total_loss}-\ref{eq:empty_space_loss}) does not specify the value the network should take anywhere other the zero-level set. Depending on initialization, the network can take negative values outside the surface. The fact that we want the network function to be positive outside the surface and negative inside is expressed in the loss function only through the term for the normals and, hence, in a local fashion. Early in training, this normal term will lead to the network taking positive values outside the surface but only close to the boundary. Far from the surface the network function continue to be negative. This creates the outer shell we see in the figure. Once the shell is created, the eikonal term $L_{eik}$ (equation \ref{eq:eikonal_loss}) and the term that penalizes small values of the network function $L_{es}$ (equation \ref{eq:empty_space_loss}) lead to a local minimum.

\begin{table*}[t]
    \centering
    \begin{tabular}{|l||c|c|c|c|c|c|}
        \hline
        \multirow{3}{*}{Shape} & \multicolumn{6}{c|}{Mean Chamfer Distance $\times 10^3$ ($\downarrow$)} \\
        \cline{2-7}
        & \multicolumn{3}{c|}{Over whole surface} & \multicolumn{3}{c|}{Inside interaction area} \\
        \cline{2-7}
        & Ours & Naive & Simple Mesh & Ours & Naive & Simple Mesh \\
        \hline
        Dining chair & \textbf{7.256} & 20.545 & 8.260 & \textbf{8.843} & 11.288 & 10.703 \\
        Chair        & \textbf{8.498} & 30.620 & 8.650 & \textbf{8.741} & 19.072 & 15.344 \\
	    Armchair     & \textbf{14.152} & 19.433 & 14.311 & \textbf{5.085} & 10.942 & 23.654 \\
	    Sofa         & \textbf{11.160} & 18.268 & 11.899 & \textbf{4.477} & 8.623 & 23.940 \\
	    Vase         & \textbf{9.063} & 15.565 & 10.345 & \textbf{10.477} & 15.713 & 16.725 \\
        \hline \hline
        \textit{Average} & \textbf{10.026} & 20.8862 & 10.693 & \textbf{7.525} & 13.128 & 18.073 \\
        \hline
    \end{tabular}
    \caption{Comparison of our editing method with and without model samples (Ours and Naive, respectively) and direct mesh editing on a mesh with equivalent size (Simple Mesh). Chamfer distances are computed with 100000 points. The mean for each shape is taken over 10 independent edits.}
    \label{tab:shapenet_results}
\end{table*}

The training that proceeds can be regarded through a variational point of view. $L_{eik}$ acts as a constraint on the network function forcing it to be an SDF, which means that the outer shell created will change as a surface in a continuous manner during gradient descent. The minimization of $L_{es}$ requires that the area of the outer shell decreases and so the shell shrinks. When the outer shell is close enough to the surface of the shape which the network is trained to represent the shell cannot shrink further because of the network's limited capacity. Between the shell's surface and the shape's surface there exists a discontinuity in the SDF's gradient. This discontinuity, of course, exists even when the shell is far from the shape, however when it gets close the network's gradient is required to change too abruptly which is not possible for its size. Thus, the training is stuck at a local minimum.

We notice that larger networks sizes and/or more complex shapes help to avoid the issue. The solution we opted for is pre-training the network. That is, before starting the actual training we execute 100 iterations using the following loss function:

\begin{equation}
    L_{pre} = \mathbb{E}_{q} \big\{ \big| f_{\theta}(\bm{x}) - \| \bm{x} \| \big| \big\}
\end{equation}

where $\theta$ is the parameter vector, $f_{\theta}$ is the network function, and $q$ is the uniform distribution in a bounding box. This leads to the network function being positive which avoids the creation of the shell.

\section{Additional PDF Estimation Results}

In the main paper, we provided PDF Estimation results only for the bunny and the sphere in the main paper (Section \ref{sec:PDFEstimation}). In Figure \ref{fig:supp_pdf_results}, we provide them for all the shapes in the dataset. Here, as well, it is clear the our proposed algorithm produces a more uniform distribution than the naive approach.

\section{Additional Surface Editing Comparisons}

We also run the surface editing comparison experiment for five shapes from the ShapeNet dataset \cite{shapenet} shown in Figure \ref{fig:shapenet_shapes}.
Since the ShapeNet meshes do not have adequately high resolution to capture the desired edits accurately and in order to have a fair comparison between the approaches, we consider as ground truth a high resolution mesh constructed by Marching Cubes \cite{marching_cubes}. This is similar to the process that we adopted for the Sphere and Torus shapes in the experimental comparisons of the main paper.
We follow the same protocol of experimental comparisons as in Section~\ref{sec:MeshComp} of the main paper and the results are reported in Table~\ref{tab:shapenet_results}. We observe that, once again, our method outperforms the compared approaches consistently for all shapes and with respect to both metrics used (i.e.\ Mean Chamfer Distance over the whole surface and inside the interaction area).

{\small
\printbibliography
}

\end{refsection}

\end{appendices}

\end{document}